\documentstyle[preprint,eqsecnum,aps]{revtex} %
\draft
\begin{document}                 
\tighten
\title{Gravitational waves from inspiralling compact binaries:\\
Energy loss and waveform to second--post-Newtonian order}
\author{Luc Blanchet}
\address{D\'epartement d'Astrophysique Relativiste et de Cosmologie,\\
Centre National de la Recherche Scientifique (UPR 176),\\
Observatoire de Paris, 92195 Meudon Cedex, France}
\author{Thibault Damour}
\address{Institut des Hautes Etudes Scientifiques,\\
91440 Bures sur Yvette, France, and \\
D\'epartement d'Astrophysique Relativiste et de Cosmologie,\\
Centre National de la Recherche Scientifique (UPR 176),\\
Observatoire de Paris, 92195 Meudon Cedex, France}
\author{Bala R. Iyer}
\address{Raman Research Institute,\\
Bangalore 560080, India}
\date{\today}
\maketitle
\begin{abstract}
Gravitational waves generated by inspiralling compact binaries are
investigated to the second--post-Newtonian (2PN) approximation of general
relativity.  Using a recently developed 2PN-accurate wave generation
formalism, we compute the gravitational waveform and associated energy
loss rate from a binary system of point-masses moving on a quasi-circular
orbit. The crucial new input is our computation of the
2PN-accurate ``source'' quadrupole moment of the binary.  Tails in both
the waveform and energy loss rate at infinity are explicitly computed.
Gravitational radiation reaction effects on the orbital frequency and
phase of the binary are deduced from the energy loss.  In the limiting
case of a very small mass ratio between the two bodies we recover the
results obtained by black hole perturbation methods.  We find that
finite mass ratio effects are very significant as they increase the 2PN
contribution to the phase by up to 52\%.  The results of this paper
should be of use when deciphering the signals observed by the future
LIGO/VIRGO network of gravitational-wave detectors.
\end{abstract}

\section{Introduction}
\label{sec:1}

The LIGO/VIRGO network of kilometer-size interferometric detectors of
gravitational waves is expected to be in operation by the turn of the
century \cite{LIGO,VIRGO} (see \cite{Th94} for a recent review).  The
most promising targets for this network are the gravitational waves
emitted during the radiation-reaction-driven inspiral of binary systems
of compact objects (neutron stars or black holes).  Crucial to the
successful detection and deciphering of such waves will be the
availability of accurate theoretical templates for the inspiral
gravitational waveforms \cite{3mn,FCh93,CF94}.  Much theoretical effort
is currently being spent on developing improved formalisms tackling the
generation of gravitational waves by general material sources and/or on
applying existing formalisms to the explicit computation of increasingly
accurate inspiral waveforms.  Among the existing generation formalisms,
the one proposed by us \cite{BD89,DI91a,BD92} can, in principle, be
developed to an arbitrarily high accuracy.  Recent work by one of us
\cite{B94} has succeeded in pushing its accuracy to the second
post-Newtonian (2PN) level, i.e.  in deriving general expressions for
the asymptotic gravitational waveform, as a functional of the matter
distribution in the source, which take into account all contributions of
fractional order $\varepsilon^4$ beyond the leading (``quadrupole
formula'') term.  Here $\varepsilon \sim v/c \sim (Gm/rc^2)^{1/2}$
denotes the small parameter entering the post-Newtonian expansion
appropriate to the description of slowly moving, weakly stressed, weakly
self-gravitating systems.

The object of the present paper is to apply the 2PN-accurate generation
formalism of Refs.~\cite{BD89,DI91a,BD92,B94} to the specific case of an
inspiralling compact binary. This is a non trivial task as the end results
of Ref.~\cite{B94} contain some complicated three-dimensional integrals
which are mathematically defined by a procedure of analytic continuation.
Note that this contrasts with the end results of our previous 1.5PN-accurate
generation formalism \cite{BD89,DI91a,BD92} which contained only integrals
extending over the (compact) support of the material source. The higher
complexity of the 2PN-level is due to the appearance of terms associated
with the cubic nonlinearities of Einstein's equations. We shall show
below how to explicitly compute these terms in the case of binary systems.

Theoretical waveforms such as the one computed below are useful to
define parametrized ``chirp'' templates to be cross correlated with the
outputs of the LIGO/VIRGO interferometric detectors. For this technique
to be successful, the templates must remain in phase with the exact
general-relativistic waveform as long as possible. The phase of the signal
is determined by the rate of change of the orbital period resulting from
gravitational radiation reaction effects.  Following the usual heuristic
approach (which has been validated in detail at the leading order
\cite{DD81a}), the effect of gravitational radiation reaction on the
orbital period can be computed from the losses of energy (and angular
momentum) \cite{N1} at infinity.  We shall therefore pay special
attention to the computation of the 2PN-accurate energy loss from a
(quasi-) circular compact binary which is obtained here for the first
time, together with the resulting orbital phasing of the binary.

Several investigations have recently focused on the computation of the
energy loss and waveform in the case of a very small mass ratio between
the two bodies \cite{P93,CFPS93,TNaka94,TSasa94}. Notably the energy
loss has been computed in this case both numerically \cite{TNaka94} and
analytically \cite{TSasa94} up to an order going well beyond the
2PN-level.  The test-mass limit of our result agrees with the 2PN
truncation of the results of Refs.~\cite{TNaka94,TSasa94}.  However, we
find that finite mass effects change very significantly the 2PN
numerical contribution to the accumulated orbital phase.  Our main
results, completed by the contributions due to the spins of the orbiting
bodies, have been briefly reported in \cite{BDIWW94}.  Note that our
formula for the energy loss has been confirmed by an independent
derivation based on a different, albeit less rigorous, method \cite{WW94}.

The organization of the paper is the following.  In Section~\ref{sec:2}
we write down the results of the 2PN generation formalism in a form
convenient for the application to inspiralling binaries and we outline our
strategy. In Section~\ref{sec:3} we compute the 2PN-accurate ``source''
moments of mass-type (especially the quadrupole moment $\ell =2$) in the
case of a binary system made of two point masses. In particular a crucial
cubically nonlinear term is obtained in the subsection~\ref{sec:3.3}.
The expressions for all the relevant source moments are given in
Section~\ref{sec:4} which deals with the explicit computation of the
2PN-accurate waveform and energy loss rate (including relevant tails).
The instantaneous orbital phase of the binary is computed at the end of
Section~\ref{sec:4}. Technical details are relegated to several appendices:
the conserved mass monopole and dipole moments are considered in
Appendix~\ref{sec:apa};  Appendix~\ref{sec:apb} presents an alternative
derivation of the cubically nonlinear contribution to the quadrupole
moment, which is valid for N-body systems;  and Appendix~\ref{sec:apc}
is a compendium of various formulas for moments.

\section{Summary of the 2PN-accurate generation formalism}
\label{sec:2}

 Let us first recall that in a suitable ``radiative'' coordinate system
$X^\mu = (cT,X^i)$ the metric coefficients, say $G_{\alpha\beta}
(X^\gamma)$, describing the gravitational field outside an isolated
system admit an asymptotic expansion in powers of $R^{-1}$, when
$R=|{\bf X}| \to \infty$ with $T-R/c$ and ${\bf N}\equiv {\bf X}/R$
being fixed (``future null infinity''). The transverse-traceless ($TT$)
projection of the deviation of $G_{\alpha\beta} (X^\gamma)$ from the
flat metric (signature $-1,+1,+1,+1$) defines the
asymptotic waveform $h^{TT}_{km} \equiv (G_{km} (X) -\delta_{km})^{TT}$
(latin indices $i,j,k,m$\dots range from 1 to 3).
The $1/R$ part of $h^{TT}_{km}$ can be uniquely decomposed into
multipoles:
\begin{eqnarray}
 h^{TT}_{km} ({\bf X},T) =&& {4G\over c^2R} {\cal P}_{ijkm} ({\bf N})
 \sum^\infty_{\ell =2} {1\over c^\ell \ell !} \biggl\{ N_{L-2} U_{ijL-2}
 (T_R) \nonumber \\
 &&- {2\ell \over (\ell +1)c} N_{aL-2}
 \varepsilon_{ab(i} V_{j)bL-2}  (T_R)\biggr\} +
 O \left( {1\over R^2}\right)\ . \label{eq:2.1}
\end{eqnarray}
The ``radiative'' multipole moments $U_L$ and $V_L$ (defined for $\ell
\geq 2$) denote some functions of the retarded time $T_R\equiv T-R/c$
taking values in the set of symmetric trace-free (STF) three-dimensional
cartesian tensors of order $\ell$.  Here $L\equiv i_1\dots i_\ell$
denotes a spatial multi-index of order $\ell$, $N_{L-2} \equiv N_{i_1}
\dots N_{i_{\ell-2}}$, $X_{(ij)} \equiv {1\over 2} (X_{ij} +X_{ji})$ and
\begin{equation}
 {\cal P}_{ijkm} ({\bf N}) = (\delta_{ik} -N_iN_k)(\delta_{jm} -N_jN_m)
 -{1\over 2} ( \delta_{ij} -N_iN_j) (\delta_{km}-N_kN_m)\ .
 \label{eq:2.2}
\end{equation}
[For the convenience of the reader we summarize our notation in
Ref.~\cite{N}.] As indicated in Eq.~(\ref{eq:2.1}), for slowly moving systems
the multipole order is correlated with the post-Newtonian order. The
coefficients in Eq.~(\ref{eq:2.1}) have been chosen so that the moments
$U_L$ and $V_L$ reduce, in the non-relativistic limit $c\to +\infty$
(or $\varepsilon \to 0$), to the $\ell$-th time-derivatives of the usual
Newtonian mass-type and current-type moments of the source. At the 2PN
approximation, i.e. when retaining all terms of fractional order
$\varepsilon^4 \sim c^{-4}$ with respect to the leading (Newtonian
quadrupole) result, the waveform (\ref{eq:2.1}) reads
\begin{eqnarray}
 h^{TT}_{km} = {2G\over c^4R} {\cal P}_{ijkm} \biggl\{  U_{ij}
  &+& {1\over c} \left[ {1\over 3} N_a U_{ija} + {4\over 3}
   \varepsilon_{ab(i} V_{j)a} N_b \right] \nonumber \\
  &+& {1\over c^2} \left[ {1\over 12} N_{ab} U_{ijab} + {1\over 2}
   \varepsilon_{ab(i} V_{j)ac} N_{bc} \right] \nonumber \\
  &+& {1\over c^3} \left[ {1\over 60} N_{abc} U_{ijabc} + {2\over 15}
   \varepsilon_{ab(i} V_{j)acd} N_{bcd} \right] \nonumber \\
  &+& {1\over c^4} \left[ {1\over 360} N_{abcd} U_{ijabcd} + {1\over 36}
   \varepsilon_{ab(i} V_{j)acde} N_{bcde} \right] + O(\varepsilon^5)
   \biggr\}\ .\label{eq:2.3}
\end{eqnarray}
 The rate of decrease of the Bondi energy $E_B$ with respect to the
retarded time $T_R \equiv T -R/c$ is related to the waveform by
\begin{equation}
  {dE_B\over dT_R} = - {c^3\over 32\pi G} \int \left( {\partial
   h^{TT}_{ij} \over \partial T_R} \right)^2 R^2 d\Omega ({\bf N})\ .
     \label{eq:2.4}
\end{equation}
At the 2PN approximation this yields (with $U^{(n)}\equiv d^nU/dT_R^n)$
\begin{eqnarray}
 {dE_B\over dT_R}=-{G\over c^5}\biggl\{ {1\over 5} U^{(1)}_{ij} U^{(1)}_{ij}
 &+&{1\over c^2} \left[ {1\over 189} U^{(1)}_{ijk} U^{(1)}_{ijk}
 + {16\over 45} V^{(1)}_{ij} V^{(1)}_{ij}\right] \nonumber \\
 &+&{1\over c^4} \left[ {1\over 9072} U^{(1)}_{ijkm} U^{(1)}_{ijkm}
 + {1\over 84} V^{(1)}_{ijk} V^{(1)}_{ijk}\right]+O(\varepsilon^6)\biggr\}\ .
 \label{eq:2.5}
\end{eqnarray}

A 2PN-accurate gravitational wave generation formalism is a method allowing
one to compute the radiative moments entering Eqs.~(\ref{eq:2.3}) and
(\ref{eq:2.5}) in terms of the source variables with an accuracy
sufficient for obtaining the waveform with fractional accuracy $1/c^4$.
The latter requirement implies, in view of Eq.~(\ref{eq:2.3}), that one
should (at a minimum) compute:  the mass-type quadrupole radiative
moment $U_{i_1i_2}$ with $1/c^4$ accuracy;  the mass-type radiative
octupole $U_{i_1i_2i_3}$ and the current-type radiative quadrupole
$V_{i_1i_2}$ with $1/c^3$ accuracy;  $U_{i_1i_2i_3i_4}$ and
$V_{i_1i_2i_3}$ with $1/c^2$ accuracy;  $U_{i_1i_2i_3i_4i_5}$ and
$V_{i_1i_2i_3i_4}$ with $1/c$ accuracy;  and $U_{i_1i_2i_3i_4i_5i_6}$
and $V_{i_1i_2i_3i_4i_5}$ with the Newtonian accuracy.  Note that these
requirements are relaxed if one is only interested in getting the energy
loss rate with 2PN-accuracy.  In that case, Eq.~(\ref{eq:2.5}) shows
that one still needs $U_{i_1i_2}$ with $1/c^4$ accuracy, but that it is
enough to compute $U_{i_1i_2i_3}$ and $V_{i_1i_2}$ with $1/c^2$
accuracy, and $U_{i_1i_2i_3i_4}$ and $V_{i_1i_2i_3}$ with Newtonian
accuracy.

 In our generation formalism, the link between the radiative multipoles
$U_L$ and $V_L$ and the dynamical state of the material source is
obtained in several steps involving as intermediate object a certain
vacuum ``canonical" metric $g^{\rm can}_{\mu\nu} (x^\lambda_{\rm can})$
expressed in terms of some ``canonical'' multipoles $M_L$ and $S_L$
(alternatively referred to as algorithmic moments in \cite{DI91a}).
On the one hand, the matching of $g^{\rm can}_{\mu\nu} (x^\lambda_{\rm
can})$ to a (PN-expanded) near-zone solution of the inhomogeneous Einstein
equations allows one to compute $M_L$ and $S_L$ in terms of some suitably
defined ``source'' multipoles $I_L$[source], $J_L$[source]. On the other
hand, the computation of nonlinear effects in the wave zone allows one
to compute $U_L$ and $V_L$ as functionals of $M_L$ and $S_L$. The final
result for the 2PN-accurate generation formalism reads (when working in an
initially mass-centred coordinate system, i.e. such that the canonical
mass dipole $M_i$ vanishes for all times)
\begin{mathletters}
\label{eq:2.6}
\begin{eqnarray}
 U_{ij} (T_R) &=& I^{(2)}_{ij} (T_R) + {2Gm\over c^3} \int^{+\infty}_0 d\tau
  \left[ \ln \left({\tau\over 2b}\right) + {11\over 12}\right] I^{(4)}_{ij}
   (T_R-\tau) + O(\varepsilon^5)\ , \label{eq:2.6a} \\
 U_{ijk} (T_R) &=& I^{(3)}_{ijk} (T_R) + {2Gm\over c^3} \int^{+\infty}_0 d\tau
  \left[ \ln \left({\tau\over 2b}\right) + {97\over 60}\right] I^{(5)}_{ijk}
   (T_R-\tau) + O(\varepsilon^5)\ , \label{eq:2.6b} \\
 V_{ij} (T_R) &=& J^{(2)}_{ij} (T_R) + {2Gm\over c^3} \int^{+\infty}_0 d\tau
  \left[ \ln \left({\tau\over 2b}\right) + {7\over 6}\right] J^{(4)}_{ij}
   (T_R-\tau) + O(\varepsilon^4)\ , \label{eq:2.6c}
\end{eqnarray}
\end{mathletters}
for the moments that need to be known beyond the 1PN accuracy, and
\begin{mathletters}
\label{eq:2.7}
\begin{eqnarray}
 U_L (T_R) &=& I^{(\ell)}_L (T_R) + O(\varepsilon^3)\ , \label{eq:2.7a}\\
 V_L (T_R) &=& J^{(\ell)}_L (T_R) + O(\varepsilon^3)\ , \label{eq:2.7b}
\end{eqnarray}
\end{mathletters}
for the other ones. Eqs.~(\ref{eq:2.6}) involve some integrals which are
associated with tails; these integrals have in front of them the total
mass-energy $m$ of the source, and contain a quantity $b$ which is an
arbitrary constant (with the dimension of time) parametrizing a certain
freedom in the construction of the radiative coordinate system $(T,{\bf
X})$.  More precisely, the link between the (Bondi-type) radiative
coordinates $X^\mu =(cT, X^i)$ and the (harmonic) canonical coordinates
$x^\mu_{\rm can} = (ct_{\rm can}, x^i_{\rm can})$ reads
\begin{equation}
 T_R = t_{\rm can} - {r_{\rm can}\over c} -
 {2Gm\over c^3} \ln \left( {r_{\rm can}\over cb}\right) + O(\varepsilon^5)
 + O\left( 1/r^2_{\rm can}\right)  \ . \label{eq:2.8}
\end{equation}

 Except for the computation of $U_{ij}$ which requires the knowledge of
the mass quadrupole source moment $I_{ij}$ with 2PN accuracy, the
computation of the other multipole contributions to the waveform can be
obtained from 1PN-accurate expressions of the mass-type and current-type
source moments which have
been obtained for all values of $\ell$ in Refs.~\cite{BD89} and
\cite{DI91a} respectively, as explicit integrals extending only on the
compact support of the material source.  [Note that there are no $1/c^3$
contributions in the {\it source} moments.] Let us illustrate the
structure of the 1PN results by quoting the simpler 1PN mass-type source
moments:
\begin{eqnarray}
 I_L(t) &=& \int d^3x \left[ \hat x_L \sigma (t,{\bf x}) +
  {\hat x_L {\bf x}^2 \over 2(2\ell+3)c^2}
  {\partial^2\sigma (t,{\bf x})\over \partial t^2}\right. \nonumber\\
  &&\qquad \left.- {4(2\ell+1)\hat x_{iL}\over (\ell+1)(2\ell+3)c^2}
  {\partial\sigma_i (t, {\bf x})\over \partial t} \right]
   + O(\varepsilon^4)\ .\label{eq:2.9}
\end{eqnarray}
The (compact support) matter densities appearing in Eq.~(\ref{eq:2.9}),
and their generalizations discussed below, are defined from the
contravariant components (in the harmonic, ``source'' coordinate system
$x^\mu$) of the material stress-energy tensor $T^{\mu\nu}$ as
\begin{mathletters}
\label{eq:2.10}
\begin{eqnarray}
 \sigma (t,{\bf x}) &\equiv& {T^{00} (t,{\bf x}) + T^{ss} (t,{\bf x})
   \over c^2}\ , \label{eq:2.10a}\\
 \sigma_i (t,{\bf x}) &\equiv& {T^{0i} (t,{\bf x}) \over c}\ ,
         \label{eq:2.10b}\\
 \sigma_{ij} (t,{\bf x}) &\equiv& T^{ij} (t,{\bf x}) \ .
         \label{eq:2.10c}
\end{eqnarray}
\end{mathletters}
The powers of $c$ introduced in Eqs.~(\ref{eq:2.10}) are such that the
$\sigma$'s have a finite non zero limit as $1/c\to 0$. One
associates to the matter densities (\ref{eq:2.10}) some  Newtonian-like
potentials, say
\begin{mathletters}
\label{eq:2.11}
\begin{eqnarray}
 U({\bf x},t) &=& G \int {d^3{\bf x}'\over |{\bf x}-{\bf x}'|} \sigma
 ({\bf x}',t)\ , \label{eq:2.11a}\\
 X({\bf x},t) &=& G \int d^3{\bf x}'|{\bf x}-{\bf x}'| \sigma
 ({\bf x}',t)\ , \label{eq:2.11b}\\
 U_i({\bf x},t) &=& G \int {d^3{\bf x}'\over |{\bf x}-{\bf x}'|} \sigma_i
 ({\bf x}',t)\ , \label{eq:2.11c}\\
 P_{ij}({\bf x},t) &=& G \int {d^3{\bf x}'\over |{\bf x}-{\bf x}'|}
  \left[ \sigma_{ij} + {1\over 4\pi G} \left( \partial_i U\partial_j U
 - {1\over 2} \delta_{ij} \partial_k U\partial_k U\right) \right]
  ({\bf x}',t)\ . \label{eq:2.11d}
\end{eqnarray}
\end{mathletters}
[Note the equation satisfied by the $X$ potential: $\Delta X=2U$.] The
2PN-accurate source moment $I_L(t)$ has been obtained in Ref.~\cite{B94}
and expressed in terms of the matter densities $\sigma$, $\sigma_i$,
$\sigma_{ij}$, the potentials $U$, $U_i$, $P_{ij}$ and the trace
$P\equiv P_{ss}$.  The result (see Eq.~(4.21) of Ref.~\cite{B94}) is
\begin{eqnarray}
 I_L(t) &=& {\rm FP}_{B=0}\int d^3 {\bf x}|{\bf x}|^B
 \biggl\{ \hat x_L \biggl[\sigma + {4\over c^4}(\sigma_{ii}U -\sigma P)
 \biggr] + {|{\bf x}|^2\hat x_L\over 2c^2(2\ell+3)} \partial^2_t
\sigma\nonumber\\
 && -{4(2\ell+1)\hat x_{iL}\over c^2(\ell+1)(2\ell+3)} \partial_t
   \left[ \left( 1 +{4U\over c^2} \right) \sigma_i  +
  {1\over \pi Gc^2} \left( \partial_k U[\partial_i U_k -\partial_k U_i]
  + {3\over 4} \partial_t U \partial_i U \right) \right] \nonumber \\
 && +{|{\bf x}|^4\hat x_L\over 8c^4(2\ell+3)(2\ell+5)} \partial^4_t\sigma
   - {2(2\ell+1)|{\bf x}|^2\hat x_{iL}\over c^4(\ell+1)(2\ell+3)(2\ell+5)}
   \partial^3_t\sigma_i  \nonumber\\
 && + {2(2\ell+1)\over c^4(\ell+1)(\ell+2)(2\ell+5)} \hat x_{ijL}
   \partial_t^2 \left[ \sigma_{ij} + {1\over 4\pi G} \partial_i U
   \partial_j U \right] \nonumber\\
 &&  + {1\over \pi Gc^4}\hat x_L \biggl[ -P_{ij} \partial_{ij} U
  - 2 U_i\partial_t\partial_i U + 2\partial_i U_j \partial_j U_i
  - {3\over 2} (\partial_t U)^2 -U\partial_t^2 U\biggr] \biggr\}
   + O(\varepsilon^5)\ . \label{eq:2.12}
\end{eqnarray}
This expression obviously reduces to Eq.~(\ref{eq:2.9}) at the 1PN-level.
It was also shown in Eq.~(4.13) of Ref.~\cite{B94} that the current moment
$J_L$, which needs only 1PN accuracy, can be written as
\begin{eqnarray}
 J_L(t) &=& {\rm FP}_{B=0}\,\varepsilon_{ab<i_\ell}
   \int d^3 {\bf x}|{\bf x}|^B
   \biggl\{ \hat x_{L-1>a}\left( 1+{4\over c^2}U\right) \sigma_b
  + {|{\bf x}|^2\hat x_{L-1>a}\over
   2c^2 (2\ell+3)} \partial^2_t \sigma_b \nonumber \\
 && \qquad\qquad + {1\over \pi Gc^2} \hat x_{L-1>a}
  \left[ \partial_k U(\partial_b U_k -\partial_k U_b)
  + {3\over 4} \partial_t U \partial_b U \right] \nonumber \\
 &&\qquad\qquad - {(2\ell +1)\hat x_{L-1>ac}\over c^2(\ell+2)(2\ell+3)}
 \partial_{t}\left[ \sigma_{bc} +{1\over 4\pi G}  \partial_b U\partial_c
   U \right] \biggr\} + O(\varepsilon^4)\ .  \label{eq:2.13}
\end{eqnarray}
This form is equivalent to the result previously derived in
Ref.~\cite{DI91a} (Eq.~(5.18) there).  See below for comments on the
symbol ${\rm FP}_{B=0}$ in front of Eqs.~(\ref{eq:2.12})-(\ref{eq:2.13}).

In this paper, it will be convenient to split the potential
$P_{ij}$ of Eq.~(\ref{eq:2.11d}) into a compact-source potential
$U_{ij}$ and a nonlinear potential $W_{ij}$ according to
\begin{mathletters}
\label{eq:2.14}
\begin{equation}
 P_{ij} = U_{ij} -W_{ij} +{1\over 2} \delta_{ij} W_{ss}\ ,\label{eq:2.14c}
\end{equation}
where $U_{ij}$ and $W_{ij}$ are defined by
\begin{eqnarray}
 U_{ij} ({\bf x},t) &=& G \int {d^3{\bf x}'\over |{\bf x}-{\bf x}'|}
 \sigma_{ij} ({\bf x}',t)\ , \label{eq:2.14a}\\
 W_{ij} ({\bf x},t) &=& - {1\over 4\pi} \int
   {d^3{\bf x}'\over |{\bf x}-{\bf x}'|} [\partial_i U \partial_j U]
  ({\bf x}',t)\ . \label{eq:2.14b}
\end{eqnarray}
\end{mathletters}
[We have $\Delta W_{ij} =\partial_i U\partial_j U$.] The compact-source
potential $U_{ij}$ should not be confused with the previously
defined radiative quadrupole moment $U_{ij}$. Also note that the
notation $W_{ij}$
is the same as used in Ref.~\cite{B94}
to denote a different potential (which is a retarded version of the
potential $P_{ij}$) but which will not be used in this paper.
Let us note for future reference that the trace of $W_{ij}$ as defined
here satisfies
\begin{mathletters}
\label{eq:2.15}
\begin{eqnarray}
 W_{ss} &=& {1\over 2} U^2 - \Phi \ , \label{eq:2.15a} \\
 \Phi ({\bf x},t) &=& G \int {d^3{\bf x}'\over |{\bf x}-{\bf x}'|}
 [\sigma U] ({\bf x}',t)\ . \label{eq:2.15b}
\end{eqnarray}
\end{mathletters}
The potentials introduced above are connected by the approximate
differential identities
\begin{mathletters}
\label{eq:2.16}
\begin{eqnarray}
 \partial_t U + \partial_i U_i &=& O(\varepsilon^2)\ , \label{eq:2.16a}\\
 \partial_t U_i + \partial_j U_{ij} &=& \partial_j \left( W_{ij} -
 {1\over 2} \delta_{ij} W_{ss}\right) + O(\varepsilon^2)\ . \label{eq:2.16b}
\end{eqnarray}
\end{mathletters}
We replace in Eq.~(\ref{eq:2.12}) the potential $P_{ij}$ by
Eqs.~(\ref{eq:2.14}) above, and $\partial_t U$ by the spatial derivative
$-\partial_i U_i$. It is also convenient to perform an integration by
parts, using $\partial_k U\partial_k U_i \equiv {1\over 2} [\Delta (UU_i)
-U\Delta U_i - U_i\Delta U]$, which is justified by a reasoning
similar to the ones followed in Sect.~IV of Ref.~\cite{B94}. This leads
to the form we shall use as starting point in this work:
\begin{eqnarray}
 I_L (t) &=& {\rm FP}_{B=0} \int d^3{\bf x} |{\bf x}|^B \left\{ \hat x_L
 \left[\sigma -{4\over c^4}\sigma U_{ss} +{4\over c^4} U\sigma_{ss}\right]
 + {|{\bf x}|^2\hat x_L\over 2c^2(2\ell+3)} \partial^2_t \sigma
    \right.\nonumber\\
 &&- {4(2\ell+1) \hat x_{iL}\over c^2(\ell+1)(2\ell+3)} \partial_t
 \left[ \left( 1+{2U\over c^2}\right) \sigma_i -{2U_i\over c^2}\sigma
 + {1\over \pi Gc^2} \left( \partial_jU \partial_i U_j - {3\over 4}
  \partial_i U\partial_j U_j\right) \right]\nonumber \\
 &&+ {|{\bf x}|^4 \hat x_L\over 8c^4(2\ell+3)(2\ell+5)} \partial^4_t
 \sigma -{2(2\ell+1)|{\bf x}|^2\hat x_{iL}\over c^4(\ell+1)(2\ell+3)(2\ell+5)}
  \partial^3_t \sigma_i \nonumber \\
 &&+ {2(2\ell+1) \hat x_{ijL}\over c^4(\ell+1)(\ell+2)(2\ell+5)}
\partial^2_t \left[\sigma_{ij}+{1\over 4\pi G} \partial_i U\partial_j U\right]
\nonumber \\
 &&\left.+ {\hat x_L\over \pi Gc^4} \left[ 2U_i \partial_{ij} U_j
 - U_{ij} \partial_{ij} U - {1\over 2}(\partial_i U_i)^2
  +2\partial_i U_j \partial_j U_i - {1\over 2} \partial^2_t
 (U^2) + W_{ij} \partial_{ij} U \right] \right\}\ . \label{eq:2.17}
\end{eqnarray}
For simplicity, we henceforth drop most of the post-Newtonian error
terms as they are usually evident from the context.

 The symbol ${\rm FP}_{B=0}$ in Eq.~(\ref{eq:2.17}) stands for ``Finite
Part at $B=0$'' and denotes a mathematically well-defined operation of
analytic continuation. Let us recall its precise meaning (see
\cite{BD86,BD88} for details and proofs of the applicability of such
a definition): one considers separately two functions of one complex
variable $B$ defined by the integrals: $I_1(B)\equiv \int_{V_1} d^3 {\bf x}
|{\bf x}|^B f({\bf x})$, $I_2 (B) \equiv \int_{V_2} d^3 {\bf x}|{\bf x}|^B
f({\bf x})$, where $V_1$ is, say, the ball $0\leq |{\bf x}| \leq r_0$
and $V_2$ the complementary domain: $|{\bf x}| > r_0$. If the function
$f({\bf x})$ is, say, continuous in $I\!\!R^3$ and has, at most, a
polynomial growth $O(|{\bf x}|^p)$ when $|{\bf x}|\to \infty$, the
integral defining $I_1(B)$ is convergent if the real part of $B$ is
large enough, say $Re(B) >-3$, while the integral defining $I_2(B)$ is
convergent if $Re(B) <-p-3$. If moreover the function $f({\bf x})$ admits an
asymptotic expansion (``multipolar expansion") of the form $f({\bf x})
\sim \Sigma_{k\leq p}f_{k,L} |{\bf x}|^k \hat n^L$ as $|{\bf x}|\to
\infty$ (we do not include logarithms of $|{\bf x}|$ to simplify), the
function $I_2(B)$ can be analytically continued as a meromorphic
function in the complex $B$ plane up to arbitrarily large values of
$Re(B)$ (with possible poles on the real axis). Finally, one considers
the sum of $I_1(B)$ and of the analytic continuation of $I_2(B)$ for
values of $B$ near 0: the constant term (zeroth power of $B$) in the
Laurent expansion of $I_1(B) +{\rm Analytic\ Continuation} (I_2(B))$
around $B=0$ defines ${\rm FP}_{B=0} \int d^3{\bf x}|{\bf x}|^B f({\bf x})$.
It is easily shown that this definition is independent of the choice of
the intermediate radius $r_0$ used to split $I\!\!R^3$ in two regions.
Note that, in principle, one should introduce a length scale to
a-dimensionalize $|{\bf x}|$ before taking its $B$th power in
(\ref{eq:2.17}). This is superfluous in our case because \cite{B94} has
shown that there arise no poles at $B=0$ and therefore no associated
logarithms.

 Although the result (\ref{eq:2.17}) is mathematically well-defined
(contrary to the results of Refs.~\cite{EW75,Th80} which are
expressed in terms of undefined, divergent integrals), it is a non
trivial task to compute it explicitly in terms of the source variables
only. This will be done in the next section, in the case where the
source is a binary system of non rotating compact objects (neutron stars
or black holes). To this end, we shall heuristically represent the
stress-energy tensor of the material source as a sum of Dirac delta
functions $\delta$. More generally, the stress-energy tensor of a system
of $N$ (non rotating) compact bodies is formally given by
\begin{equation}
 T^{\mu\nu} ({\bf x},t) = \sum^N_{A=1} m_A {dy^\mu_A\over dt}
 {dy^\nu_A\over dt} {1\over \sqrt{-g}} {dt\over d\tau} \delta ({\bf x}
 - {\bf y}_A (t))\ , \label{eq:2.18}
\end{equation}
where $m_A$ denotes the (constant) Schwarzschild mass of the $A$th
compact body.
 This yields for the source variables (\ref{eq:2.10})
\begin{mathletters}
\label{eq:2.19}
\begin{eqnarray}
 \sigma ({\bf x},t) &=& \sum_{A=1}^N \mu_A (t) \left( 1+ {{\bf v}^2_A
 \over c^2} \right) \delta ({\bf x} -{\bf y}_A (t))\ , \label{eq:2.19a}\\
 \sigma_i ({\bf x},t) &=& \sum_{A=1}^N \mu_A (t) v^i_A \delta
  ({\bf x} -{\bf y}_A (t))\ , \label{eq:2.19b}\\
 \sigma_{ij} ({\bf x},t) &=& \sum_{A=1}^N \mu_A (t) v^i_A v^j_A \delta
  ({\bf x} -{\bf y}_A (t))\ , \label{eq:2.19c}
\end{eqnarray}
\end{mathletters}
where $v^i_A \equiv dy^i_A /dt$ and
\begin{mathletters}
\label{eq:2.20}
\begin{eqnarray}
 \mu_A (t)&=&m_A [1 +(d_2)_A+(d_4)_A]\ ,\label{eq:2.20a}\\
 d_2 &\equiv& {1\over c^2} \left( {{\bf v}^2\over 2} -V \right)\ ,
   \label{eq:2.20b}\\
 d_4 &\equiv& {1\over c^4} \left( {3\over 8} {\bf v}^4 + {3\over 2}
  U{\bf v}^2 - 4 U_iv_i - 2 \Phi + {3\over 2} U^2 + 4 U_{ss} \right)\ ,
  \label{eq:2.20c}
\end{eqnarray}
\end{mathletters}
the notation $V$ being a shorthand for the combination
\begin{equation}
 V \equiv U + {1\over 2c^2}\, \partial_t^2 X\ , \label{eq:2.21}
\end{equation}
which is the potential appearing naturally in the 1PN near-zone metric
in harmonic coordinates. The subscript $A$ appearing in Eq.~(\ref{eq:2.20a})
indicates that one must replace the field point ${\bf x}$ by the position
${\bf y}_A$ of the $A$th mass point, while discarding all the ill-defined
(formally infinite) terms arising in the limit ${\bf x}\to {\bf y}_A$. For
instance,
\begin{equation}
 (U)_A = G \sum_{B\not= A} {\mu_B(t) (1+{\bf v}^2_B /c^2)\over
    |{\bf y}_A - {\bf y}_B|} \ . \label{eq:2.22}
\end{equation}
[Note that the second time derivative appearing in $V$, Eq.~(\ref{eq:2.21}),
must be explicated before making the replacement ${\bf x}\to {\bf y}_A (t)$.]
Although we do not claim to have verified it by a detailed proof, we feel
secure that the formal use of delta functions can be justified at the
2PN accuracy (and even at the 2.5PN accuracy) by combining our generation
formalism with the results of Ref.~\cite{D83a}. Indeed, the latter reference
showed (by a matching technique) that the metric generated by a system
of well-separated strongly-self-gravitating bodies was equal, up to the
3PN level, to the metric generated by a mathematically well-defined
version of delta functions. See Section~III.C of Ref.~\cite{BD92}
for a discussion, at the 1PN level, of how to combine the two
formalisms.

\section{The 2PN-accurate mass moments of a compact binary}
\label{sec:3}

 It is convenient to split the starting formula (\ref{eq:2.17}) into
three types of contributions, say
\begin{equation}
  I_L = I^{(C)}_L + I^{(Y)}_L + I^{(W)}_L\ . \label{eq:3.1}
\end{equation}
Here, $I^{(C)}_L$ (``Compact terms'') denotes the terms where, because
of the explicit presence of a source term $\sigma ({\bf x})$, $\sigma_i
({\bf x})$ or $\sigma_{ij} ({\bf x})$ (or a time derivative thereof),
the three-dimensional integral $\int d^3 {\bf x}$ extends only over the
compact support of the material source. The finite part prescription
is unnecessary for such terms.
[Note that $I^{(C)}_L$ is identical to the 2PN expansion of the exact
linearized gravity result given by Eq.~(5.33) of \cite{DI91b}.]
The ``$Y$-terms'' $I^{(Y)}_L$ (named after the quantity defined in
Eq.~(\ref{eq:3.22}) below) denote all the contributions involving the
three-dimensional integral of the product of (spatial derivatives of)
two Newtonian-like potentials, e.g.
\begin{equation}
 {\rm FP}_{B=0} \int d^3 {\bf x} |{\bf x}|^B \hat x_{ijL} \partial^2_t
  [\partial_i U\partial_j U] = {d^2\over dt^2} \left\{ {\rm FP}_{B=0} \int
 d^3 {\bf x} |{\bf x}|^B \hat x_{ijL} \partial_i U \partial_j U\right\}\ .
  \label{eq:3.2}
\end{equation}
The time derivatives appearing in the $Y$-terms have all been written
in a manner such that they can be factorized in front as total time
derivatives acting on (the finite part of) a three-dimensional integral.
Finally, $I^{(W)}_L$ denotes the only term of (\ref{eq:2.17}) involving
the three-dimensional integral of a term trilinear in source variables,
namely
\begin{equation}
 I^{(W)}_L = {1\over \pi Gc^4} {\rm FP}_{B=0} \int d^3 {\bf x}
   |{\bf x}|^B \hat x_L W_{ij} \partial_{ij} U\ , \label{eq:3.3}
\end{equation}
where we recall that $W_{ij}$, defined by (\ref{eq:2.14b}), is a bilinear
functional of $\sigma ({\bf x}',t)$.

 We shall consider in turn the three contributions to $I_L$. The
``Compact" and ``$Y$'' contributions will be evaluated in the quadrupole
case $(\ell =2)$ while the ``$W$'' contribution will be calculated for
any $\ell$. The cases $\ell =0$ and $\ell =1$ play a special role as
they do not correspond to radiative moments, but to conserved quantities.
We check in Appendix~\ref{sec:apa} the agreement with known results
for these low moments.

\subsection{The compact terms and their explicit form in the circular
two-body case}
\label{sec:3.1}

The general, $N$-extended-body expression for the ``compact'' contributions
to the 2PN mass moments reads
\begin{eqnarray}
I_L^{(C)}&=&\int d^3{\bf x}\left\{\hat{x}_L\left[\sigma
\left(1-\frac{4U_{ss}}{c^4} \right)+\frac{4U}{c^4}\sigma_{ss}+
\frac{1}{2(2\ell+3)}\left(\frac{{\bf x}^2}{c^2}\partial^2_t\sigma+
\frac{1}{4(2\ell+5)}\frac{{\bf x}^4}{c^4}\partial_t^4\sigma\right)
 \right]\right. \nonumber\\
&&-\frac{4(2\ell+1)}{(\ell+1)(2\ell+3)c^2}\hat{x}_{iL}\left[\partial_t
\left(\sigma_i (1+\frac{2U}{c^2})-\frac{2U_i}{c^2}\sigma \right)
+\frac{1}{2(2\ell+5)}\frac{{\bf x}^2}{c^2}\partial_t^3\sigma_i\right]
\nonumber\\
&&\left.+\frac{2(2\ell+1)}{(\ell+1)(\ell+2)(2\ell+5)c^4}\hat{x}_{ijL}
  \partial_t^2\sigma_{ij} \right\}\ .  \label{eq:3.4}
\end{eqnarray}
{}From Eqs.~(\ref{eq:2.19}) we obtain the corresponding point-mass form
\begin{eqnarray}
 I^{(C)}_L &=& \sum^N_{A=1} \left\{ \tilde\mu_A \left[ 1 -{4\over c^4}
   U^A_{ss} + {4\over c^4} U^A ({\bf v}_A)^2\right] \hat y^L_A
   \right. \nonumber\\
 && + {1\over 2(2\ell+3)c^2} {d^2\over dt^2} (\tilde\mu_A {\bf y}_A^2
  \hat y_A^L) + {1\over 8(2\ell+3)(2\ell+5)c^4} {d^4\over dt^4}
 (\tilde\mu_A ({\bf y}^2_A)^2 \hat y^L_A) \nonumber \\
 && -{4(2\ell+1)\over (\ell+1)(2\ell+3)c^2} {d\over dt} \left(
  \left[ \mu_A \left( 1+ {2U^A\over c^2} \right) v^i_A - {2U^A_i\over c^2}
  \tilde\mu_A \right] \hat y^{iL}_A\right)\nonumber \\
 && -{2(2\ell+1)\over (\ell+1)(2\ell+3)(2\ell+5)c^4} {d^3\over dt^3}
   (\mu_A v^i_A {\bf y}^2_A \hat y^{iL}_A ) \nonumber \\
 &&\left. +{2(2\ell+1)\over (\ell+1)(\ell+2)(2\ell+5)c^4} {d^2\over dt^2}
   (\mu_A v^i_A v^j_A \hat y^{ijL}_A ) \right\}\ , \label{eq:3.5}
\end{eqnarray}
in which we have introduced for brevity $\tilde\mu_A \equiv \mu_A
(1+{\bf v}^2_A/c^2).$  As it is written, the result (\ref{eq:3.5})
depends (at 2PN order) not only on the positions ${\bf y}_A$ and velocities
${\bf v}_A$ of the N compact bodies, but also on higher time derivatives
thereof, up to $d^4{\bf y}_A/dt^4$. To reduce the functional dependence
of $I_L^{(C)}$ to a dependence on positions and velocities only, we need to
use the post-Newtonian-expanded equations of motion of the $N$-body
system. At this juncture, we restrict ourselves to the simplest case of
a binary system evolving on a quasi-circular orbit. Consistently with
the 2PN accuracy of the generation formalism considered here, we use the
2PN truncation of the 2.5PN equations of motion of Ref.~\cite{DD81a}.
We use a (harmonic) coordinate system in which the 2PN
center of mass is at rest, at the origin. In such a coordinate system
the 2PN truncation of the equations of motion admit
exact circular periodic orbits. Using the 2PN-accurate center of mass
theorem of Ref.~\cite{DD81b}, we can express the individual
center-of-mass-frame positions of the two bodies in circular orbits
in terms of the relative position
\begin{equation}
 {\bf x} \equiv {\bf y}_1 - {\bf y}_2   \label{eq:3.6}
\end{equation}
as follows:
\begin{mathletters}
\label{eq:3.7}
\begin{eqnarray}
 {\bf y}_1 &=& [X_2 + 3\nu \gamma^2 (X_1 -X_2)] {\bf x}\ , \label{eq:3.7a} \\
 {\bf y}_2 &=& [-X_1 + 3\nu \gamma^2 (X_1 -X_2)] {\bf x}\ . \label{eq:3.7b}
\end{eqnarray}
\end{mathletters}
(Eqs.~(\ref{eq:3.7}) are obtained by setting $G_i = 0$ for
circular orbits, where $G_i$ is given by Eq.~(19) in Ref.~\cite{DD81b}.)
Here, we have denoted
\begin{equation}
 m \equiv m_1 + m_2\ , \quad X_1 \equiv {m_1\over m}\ ,\quad
    X_2 \equiv {m_2\over m} \equiv 1 -X_1\ ,\label{eq:3.8}
\end{equation}
and
\begin{eqnarray}
  \nu &\equiv & X_1 X_2 \equiv {m_1m_2\over m^2}\ , \label{eq:3.9}\\
  \gamma &\equiv & {Gm\over c^2r}\ , \label{eq:3.10}
\end{eqnarray}
with $r\equiv |{\bf x}| \equiv |{\bf y}_1 -{\bf y}_2|$ denoting the
constant (harmonic) coordinate radius of the relative orbit. One should
note in Eq.~(\ref{eq:3.7}) the absence of 1PN corrections to the usual
center-of-mass expressions. This is an accident due to the fact that
we are restricting our attention to circular orbits. [In the
non-circular case there are 1PN corrections to $X_2$ and $-X_1$ which
are proportional to ${\bf v}^2 -Gm/r$, see e.g. \cite{DD85}.]
Then the content of the 2PN equations of motion reduces to the knowledge
of the 2PN-accurate orbital frequency $\omega_{\rm 2PN}$ given by
\begin{equation}
 \omega^2_{\rm 2PN} \equiv {Gm\over r^3} \left[ 1 - (3-\nu) \gamma +
 \left( 6 + {41\over 4}\nu + \nu^2\right) \gamma^2 \right]\ ,
\label{eq:3.11}
\end{equation}
which is such that
\begin{mathletters}
\label{eq:3.12}
\begin{eqnarray}
 {\bf v} &\equiv & {d{\bf x}\over dt}\ , \label{eq:3.12a}\\
 {\bf a} &\equiv & {d{\bf v}\over dt}\equiv {d^2{\bf x}\over dt^2}
 = -\omega^2_{\rm 2PN}\, {\bf x} + O(\varepsilon^5)\ . \label{eq:3.12b}
\end{eqnarray}
\end{mathletters}
Let us note that Eqs.~(\ref{eq:3.12}) imply, as usual, that $v\equiv
|{\bf v}| = \omega_{\rm 2PN}\, r +O(\varepsilon^5)$, so that
Eq.~(\ref{eq:3.11})
implies
\begin{equation}
 {v^2\over c^2} = \gamma \left[ 1 - (3-\nu) \gamma + \left( 6
   + {41\over 4}\nu + \nu^2\right) \gamma^2 \right]\ . \label{eq:3.13}
\end{equation}
We are now in a position  to compute explicitly the ``compact'' terms
$I^{(C)}_L$, and we restrict ourselves to the quadrupole case $\ell=2$.

 Without entering into the details of the calculation of
$I^{(C)}_{ij}$, let us mention that there arise many symmetric
functions of the two masses which can be straightforwardly expressed in
terms of the total mass $m$ and of the quantity $\nu \equiv X_1X_2$ by
using the well-known fact that a symmetric polynomial in $X_1$ and
$X_2$ can be written in terms of the elementary symmetric combinations
$X_1+X_2 (\equiv 1)$ and $X_1X_2$. Useful formulas for this reduction
are:
\begin{mathletters}
\label{eq:3.14}
\begin{eqnarray}
  X^2_1 + X^2_2 &=& 1 - 2 \nu \ , \label{eq:3.14a}\\
  X^3_1 + X^3_2 &=& 1 - 3 \nu \ , \label{eq:3.14b}\\
  X^4_1 + X^4_2 &=& 1 - 4 \nu +2 \nu^2\ , \label{eq:3.14c}\\
  X^5_1 + X^5_2 &=& 1 - 5 \nu +5 \nu^2\ . \label{eq:3.14d}
\end{eqnarray}
\end{mathletters}
The computation of the quadrupole $I^{(C)}_{ij}$ for two bodies and
circular orbits is long and tedious but quite straightforward. We obtain
\begin{eqnarray}
I_{ij}^{(C)}&=& {\rm STF}_{ij}\, \nu m\left[x^{ij}-
\frac{\gamma}{42}x^{ij}(1+39\nu)+\frac{11}{21}\frac{r^2}{c^2}v^{ij}
(1-3\nu) \right.\nonumber\\
&&+\frac{\gamma^2}{1512}x^{ij}(5203-18275\nu-2785\nu^2) \nonumber\\
&&\left.+\frac{\gamma}{378}\frac{r^2}{c^2}v^{ij}(191-577\nu+109\nu^2)\right]
  \ ,\label{eq:3.15}
\end{eqnarray}
where ${\rm STF}_{ij}$ denotes the STF projection with respect to the
indices $ij$.

\subsection{The quadratically nonlinear terms}
\label{sec:3.2}

 Separating out from Eq.~(\ref{eq:2.17}) the terms involving the
three-dimensional integral of a product of (spatial derivatives of)
two Newtonian potentials, we define
\begin{eqnarray}
 I^{(Y)}_L &\equiv& {\rm FP}_{B=0} \int d^3{\bf x} |{\bf x}|^B \left\{
{\hat x_L\over \pi Gc^4} [ 2U_i \partial_{ij} U_j - U_{ij}\partial_{ij} U
       \right. \nonumber\\
 && -{1\over 2}\partial_i U_i\partial_j U_j +2\partial_i U_j\partial_j U_i
   - {1\over 2}\partial_t^2 (U^2)] \nonumber\\
 && -{4(2\ell+1)\hat x_{iL}\over (\ell+1)(2\ell+3)\pi Gc^4} \partial_t
\left(\partial_j U\partial_i U_j-{3\over 4}\partial_i U\partial_j U_j\right)
   \nonumber\\
 &&\left. +{(2\ell+1) \hat x_{ijL}\over (\ell+1)(\ell+2)(2\ell+5) 2\pi Gc^4}
   \partial_t^2 (\partial_i U\partial_j U) \right\} \ . \label{eq:3.16}
\end{eqnarray}
These terms can be deduced from the results of Ref.~\cite{DI91a}. Denoting
$r_1\equiv |{\bf x} -{\bf y}_1|$, $r_2\equiv |{\bf x} -{\bf y}_2|$,
$r_{12}\equiv |{\bf y}_1 -{\bf y}_2|$, the latter reference has introduced
the kernel
\begin{equation}
 k ({\bf x}; {\bf y}_1, {\bf y}_2) \equiv {1\over 2} \ln [(r_1 +r_2)^2
 - r_{12}^2] \label{eq:3.17}
\end{equation}
and proven that it satisfies (in the sense of distribution theory) the
identity
\begin{equation}
 {1\over r_1r_2} \equiv \Delta_x k - 2\pi \delta_{12}\ , \label{eq:3.18}
\end{equation}
where $\delta_{12}$ denotes a distribution supported on the segment
joining ${\bf y}_1$ to ${\bf y}_2$:
\begin{mathletters}
\label{eq:3.19}
\begin{equation}
 \delta_{12} ({\bf x}; {\bf y}_1, {\bf y}_2) \equiv r_{12} \int^1_0
d\alpha \, \delta ({\bf x} - {\bf y}_\alpha )\ , \label{eq:3.19a}
\end{equation}
where
\begin{equation}
 {\bf y}_\alpha \equiv (1-\alpha) {\bf y}_1 + \alpha {\bf y}_2\ .
      \label{eq:3.19b}
\end{equation}
\end{mathletters}
Moreover, the kernel $k({\bf x})$ is such that its multipolar expansion
when $|{\bf x}|\to \infty$ contains, besides a logarithmic term,
$\ln (2|{\bf x}|)$, only terms of the type $|{\bf x}|^{-\ell -2p}\hat n_L$
(with $p\in I\!\!N$). The important point, as we are going to see, is
that the multipolar expansion of $k({\bf x})$ contains no terms of the
type $|{\bf x}|^{-\ell-1}\hat n_L$, i.e. no homogeneous solutions of the
Laplace equation. [This is  the feature defining the kernel $k$ by
contrast with the kernel $g$ satisfying $\Delta_x g=(r_1r_2)^{-1}$
everywhere and containing a homogeneous piece ${1\over 2}h$ whose
distributional source is precisely the $-2\pi \delta_{12}$ term in
Eq.~(\ref{eq:3.18}); see \cite{DI91a}.]  The latter property implies that
\begin{eqnarray}
 {\rm FP}_{B=0} \int d^3{\bf x}&|{\bf x}|^B&\hat x_L \Delta_x k({\bf x})
 = {\rm FP}_{B=0} \int d^3 {\bf x}\, \Delta_x (|{\bf x}|^B \hat x_L)
 k({\bf x})\nonumber \\
 &&= {\rm FP}_{B=0} \left\{ B(B+2\ell+1) \int d^3{\bf x}|{\bf x}|^{B+\ell-2}
  \hat n_L k({\bf x}) \right\} = 0\ . \label{eq:3.20}
\end{eqnarray}
Here, the first equality is obtained by integrating by parts (the surface
term at infinity vanishing by analytic continuation from the case where
$Re(B)$ is large and negative), and the last follows from the fact that
$k({\bf x})$ is well behaved at the origin $|{\bf x}| =0$ and contains
no ``homogeneous'' terms $|{\bf x}|^{-\ell-1}\hat n_L$ in its multipolar
expansion at infinity. Indeed, going back to the definition recalled
above of analytically continued integrals, as the sum of one integral
over the ball $0\leq |{\bf x}| \leq r_0$ and one over its complement
$|{\bf x}|> r_0$, we see that the factor $B$ in front will give a zero
result, except if the integral near the origin $(0\leq |{\bf x}| \leq
\varepsilon)$ or the one near infinity ($|{\bf x}|> r_0$, with $r_0$
arbitrarily large) generates a pole at $B=0$ in the complex $B$ plane.
As $k({\bf x})$ exhibits no power-like blow up near the origin ${\bf
x}=0$ (even when the latter coincides with ${\bf y}_1$ or ${\bf
y}_2$) the integral near the origin is easily seen not to generate any
pole at $B=0$. Concerning the integral near infinity, written as
$\int^\infty_{r_0} dr\, r^{B+\ell} (\int_{S_2} d\Omega\, \hat n_Lk)$,
we see from the orthogonality of the $\hat n_L$'s over the sphere $S_2$
that only the presence of terms $\propto \hat n_L/r^{\ell+1}$ in the
multipolar expansion of $k({\bf x})$ could generate a pole through
$\int^\infty_{r_0} dr\, r^{B+\ell}r^{-\ell-1} = -B^{-1} r^B_0$.

 By combining the identity (\ref{eq:3.18}) with the result
(\ref{eq:3.20}) we conclude that
\begin{equation}
 {\rm FP}_{B=0} \int d^3 {\bf x}|{\bf x}|^B \hat x_L {1\over r_1r_2} =
 - 2\pi\, Y^L ({\bf y}_1, {\bf y}_2)\ , \label{eq:3.21}
\end{equation}
(see also Sect.~IV in Ref.~\cite{B94}) in which, following the notation
of \cite{DI91a}, we have introduced
\begin{equation}
 Y^L({\bf y}_1, {\bf y}_2) \equiv \int d^3{\bf x}\,\hat x_L \delta_{12}
 ({\bf x}; {\bf y}_1, {\bf y}_2) \equiv r_{12} \int^1_0 d\alpha\,
 y^{<L>}_\alpha \ , \label{eq:3.22}
\end{equation}
where $y^{<L>}_\alpha$ denotes the STF projection of $y^{i_1}_\alpha
\dots y^{i_\ell}_\alpha$ with $y^i_\alpha$ defined by Eq.~(\ref{eq:3.19b}).
As the dependence of $y^{<L>}_\alpha$ on $\alpha$
is polynomial it is easy to perform the integration over $\alpha$ in
(\ref{eq:3.22}) to get \cite{B94}
\begin{equation}
 Y^L ({\bf y}_1, {\bf y}_2) = {|{\bf y}_1 - {\bf y}_2|\over \ell+1}
   \sum^\ell_{p=0}
 y_1^{\langle L-P} y_2^{P\rangle} \ , \label{eq:3.23}
\end{equation}
where $y^P_2 = y^{i_1}_2\dots y^{i_p}_2$, $y^{L-P}_1 = y^{i_{p+1}}_1
\dots y^{i_\ell}_1.$ By taking derivatives of both sides of
(\ref{eq:3.21}) with respect to $y^i_1$ or $y^i_2$ and integrating over
${\bf y}_1$ and ${\bf y}_2$ after having weighted the integrand with some
source functions $\sigma_\alpha ({\bf y}_1)$, $\sigma_\beta ({\bf y}_2)$
[where $\sigma_\alpha$ denote $\sigma$, $\sigma_i$ or $\sigma_{ij}$],
one can obtain all the terms in (\ref{eq:3.16}) since these are all bilinear
in some Newtonian-like potentials. For instance, the right-hand side of
Eq.~(\ref{eq:3.2}) can be written as
\begin{equation}
 {d^2\over dt^2} \left\{ -2\pi G^2\int d^3\, {\bf y}_1 \sigma ({\bf y}_1,t)
 \int d^3 {\bf y}_2 \sigma ({\bf y}_2, t)
  \left( -{\partial\over \partial y_1^i}\right)
  \left( -{\partial\over \partial y_2^j}\right)  Y^{ijL} ({\bf y}_1,
    {\bf y}_2) \right\}\ , \label{eq:3.24}
\end{equation}
where one must be careful about the minus signs appearing in the spatial
derivatives due to $\partial r^{-1}_1 / \partial x^i = -\partial r^{-1}_1
/ \partial y^i_1$, $\partial r^{-1}_2 / \partial x^i = -\partial r^{-1}_2
/ \partial y^i_2$, and about keeping the total time derivatives factorized
in front of the whole expression. It is convenient to introduce a special
notation for the derivatives of $Y^L$ with respect to ${\bf y}_1$ and
${\bf y}_2$, say
\begin{mathletters}
\label{eq:3.25}
\begin{eqnarray}
  {}_{ij}Y^L &\equiv & {\partial\over \partial y^i_1}
   {\partial\over \partial y^j_1} Y^L  \ , \label{eq:3.25a}\\
  {}_{i}Y_j^L &\equiv & {\partial\over \partial y^i_1}
   {\partial\over \partial y^j_2} Y^L  \ , \label{eq:3.25b}\\
  Y_{ij}^L &\equiv & {\partial\over \partial y^i_2}
   {\partial\over \partial y^j_2} Y^L  \ . \label{eq:3.25c}
\end{eqnarray}
\end{mathletters}
With this notation in hand, it is easy, from the result (\ref{eq:3.21}),
to obtain the following expression for the ``$Y$-type'' contribution
to $I_L$ defined in Eq.~(\ref{eq:3.16}):
\begin{eqnarray}
I_L^{(Y)}&=&-{2G\over c^4}\int\!\!\int d^3{\bf y}_1d^3{\bf y}_2
\biggl\{2\sigma_1^s\sigma_2^kY^L_{sk}-
\sigma_1^{sk}\sigma_2Y^L_{sk}\nonumber\\
&&-\frac{1}{2}\sigma_1^s\sigma_2^k\;\;{}_sY_k^L+2\sigma_1^s\sigma_2^k\;\;
{}_kY^L_s -\frac{1}{2}\partial_t ^2(\sigma_1\sigma_2Y^L)\nonumber\\
&&-\frac{4(2\ell+1)}{(\ell+1)(2\ell+3)}\partial_t\left[\sigma_1
 \sigma_2^k\;\;{}_kY_a^{aL}
 -\frac{3}{4}\sigma_1^s\sigma_2\;\;{}_sY_a^{aL}\right] \nonumber\\
&&+\frac{2\ell+1}{2(\ell+1)(\ell+2)(2\ell+5)}\partial_t^2
 \left(\sigma_1\sigma_2\;\;{}_aY_b^{abL} \right)\biggr\}\ ,\label{eq:3.26}
\end{eqnarray}
where $\sigma_1 \equiv \sigma ({\bf y}_1,t)$, $\sigma_2 \equiv \sigma
({\bf y}_2,t)$, etc\dots.

 Finally, the point-mass limit is obtained by inserting
Eqs.~(\ref{eq:2.19}) into Eq.~(\ref{eq:3.26}). Note that, because of the
overall $c^{-4}$ factor, it is enough to use the Newtonian approximation
for the source terms [e.g. $\sigma ({\bf x},t) = \Sigma_A m_A \delta
({\bf x} -{\bf y}_A (t))+O(\varepsilon^2)$]. To increase the readability
of the result, it is convenient to introduce a shorthand for the
contractions of the derivatives (\ref{eq:3.25}) with the velocities:
\begin{mathletters}
\label{eq:3.27}
\begin{eqnarray}
 Y^L_{v_Av_C} &\equiv &v^i_A v^j_C\ Y^L_{ij}\ , \label{eq:3.27a}\\
{}_{v_Av_C}Y^L&\equiv &v^i_A v^j_C\ {}_{ij}Y^L\ , \label{eq:3.27b}\\
{}_{v_A}Y_{v_C}^L&\equiv &v^i_A v^j_C\ {}_iY_j^L\ , \label{eq:3.27c}
\end{eqnarray}
where $A,C =1,2$, as well as a mixed notation like e.g.
\begin{equation}
{}_{v_A}Y_j^L\equiv v^i_A \ {}_iY_j^L\ . \label{eq:3.27d}
\end{equation}
\end{mathletters}
 This leads, for any $\ell$, to the following expression for the
$Y$-terms
\begin{eqnarray}
I_L^{(Y)}&=&-{2G\over c^4}\sum_{A,C}m_Am_C
  \biggl\{2Y^L_{v_Av_C}-Y^L_{v_Av_A} \nonumber\\
&&-\frac{1}{2}{}\;{}_{v_A}Y^L_{v_C}+2\; {}_{v_C}Y^L_{v_A}-\frac{1}{2}
  \partial^2_t\left( Y^L\right)\nonumber\\
&&-\frac{4(2\ell+1)}{(\ell+1)(2\ell+3)}\partial_t\left[{}_{v_C}
  Y^{sL}_{s}-\frac{3}{4} {}_{v_A}Y^{sL}_{s}\right]\nonumber\\
&&+\frac{2\ell+1}{2(\ell+1)(\ell+2)(2\ell+5)}\partial_t^2
  \left[{}_aY^{abL}_{b} \right]\biggr\}\ ,\label{eq:3.28}
\end{eqnarray}
in which all functions $Y^L$ are evaluated with ${\bf y}_A$ in their first
argument, and ${\bf y}_C$ in their second one, e.g. $_{v_A}Y^L_{v_C} =
v^i_A v^j_C\ {}_iY_j^L ({\bf y}_A, {\bf y}_C)$, $Y^L_{v_Av_A} = v^i_A v^j_A
Y^L_{ij} ({\bf y}_A, {\bf y}_C)$, and in which all the
self-terms $A=C$ must be omitted. Finally, by using the expression
(\ref{eq:3.23}) for $Y^L$, a long calculation yields the following
explicit expression for the $Y$-type mass quadrupole for two bodies in
a circular orbit, and considered in the center-of-mass frame:
\begin{equation}
I_{ij}^{(Y)}=-\frac{2m\nu\gamma}{63}\, {\rm STF}_{ij} \left[\gamma
x^{ij}(55-155\nu-53\nu^2) +\frac{r^2}{c^2}v^{ij}
(-118+92\nu-10\nu^2)\right] \ .  \label{eq:3.29}
\end{equation}

\subsection{The cubically nonlinear term}
\label{sec:3.3}

 Let us now tackle the cubically nonlinear term
\begin{equation}
  I^{(W)}_L = {1\over \pi Gc^4} {\rm FP}_{B=0} \int d^3 {\bf x}
 |{\bf x}|^B \hat x^L W_{ij} \partial_{ij} U\ , \label{eq:3.30}
\end{equation}
with $W_{ij} = \Delta^{-1} (\partial_i U \partial_j U)$. We shall show
how to evaluate $I^{(W)}_L$ explicitly, for all values of $\ell$, in the
case where the source is a binary system. [Note in passing that as
(\ref{eq:3.30}) depends only on the instantaneous mass distribution of
the source our result is valid whatever be the orbit (circular or not)
of the binary.] To do this for all values of $\ell$ we need what is
essentially a generalization of the method of \cite{DI91a}, i.e. a
detailed study of some cubically nonlinear kernel. In the particular
(and most urgently needed) case of the mass quadrupole, $\ell =2$, one
can evaluate (\ref{eq:3.30}) by other means, as is shown in
Appendix~\ref{sec:apb} which succeeds in computing $I^{(W)}_{ij}$ for
N-body systems. This gives us an independent check of the results below.

 Let us define a function $W({\bf x},t)$ [which is a trilinear, nonlocal
functional of $\sigma ({\bf x}',t)$] by
\begin{mathletters}
\label{eq:3.31}
\begin{eqnarray}
 \Delta W &=& -{4\over c^4} W_{ij} \partial_{ij} U\ ,\label{eq:3.31a}\\
 W({\bf x}) &=& +{1\over \pi c^4} \int {d^3{\bf x}'\over |{\bf x}-{\bf x}'|}
  W_{ij} ({\bf x}') \partial_{ij} U({\bf x}')\ , \label{eq:3.31b}
\end{eqnarray}
\end{mathletters}
where, for brevity, we suppress the dependence on the time variable which
is the same on both sides. Note that the integral (\ref{eq:3.31b}) is a
usual, convergent integral [at infinity $W_{ij} =O(1/r)$, $\partial_{ij}
U =O (1/r^3)$].  As usual, $W$ is characterized as being the unique
solution (in the sense of distribution theory) of (\ref{eq:3.31a}) which
falls off at spatial infinity.

 Our method for computing $I_L^{(W)}$ is similar to the one we used
in Eqs.~(\ref{eq:3.17})-(\ref{eq:3.20}) above to compute $I_L^{(Y)}$
from the results of \cite{DI91a} on the kernels $k$ and $g$. Inserting
(\ref{eq:3.31a}) into (\ref{eq:3.30}) gives
\begin{equation}
 I^{(W)}_L = -{1\over 4\pi G} {\rm FP}_{B=0} \int d^3 {\bf x}|{\bf x}|^B
 \hat x^L \Delta_x W ({\bf x})\ . \label{eq:3.32}
\end{equation}
Integrating by parts the two spatial derivatives of $\Delta_x$ (the
analytic continuation from a case where $Re (B)$ is large and negative
ensuring the vanishing of any surface term at infinity), using the
formula $\Delta_x (|{\bf x}|^B \hat x^L) = B (B+2\ell +1) r^{B+\ell -2}
\hat n^L$, where $r\equiv |{\bf x}|$, $n^i\equiv x^i/r$, and writing
explicitly the definition of the analytically continued integral (in
polar coordinates: $d^3{\bf x} =r^2 dr d\Omega$) we get
\begin{eqnarray}
 I^{(W)}_L &=& -{1\over 4\pi G}{\rm FP}_{B=0} \left\{ B(B+2\ell +1) \left[
 \int^{r_0}_0 dr \int d\Omega\, r^{B+\ell} \hat n^L W\right.\right. \nonumber\\
 &&\qquad\qquad\qquad \qquad\qquad
\left. \left. + \int^\infty_{r_0} dr\int d\Omega
 \, r^{B+\ell} \hat n^L W \right] \right\}\ . \label{eq:3.33}
\end{eqnarray}
Because of the regularity of $W({\bf x})$ near the origin ${\bf x} =0$
\cite{N2},
the first integral on the
right-hand side of (\ref{eq:3.33}) will continuously depend on $B\in
I\!\!\!\!C$ near $B=0$, and will therefore not contribute to $I_L^{(W)}$
because of the explicit $B$ factor in front. We are therefore left with
(since the second integral can have at most a simple pole)
\begin{equation}
 I^{(W)}_L = -{2\ell+1\over 4\pi G}\, {\rm FP}_{B=0} \left\{ B
 \int^\infty_{r_0} dr\, r^{B+\ell} \left[ \int d \Omega\, \hat n^L
 W({\bf x}) \right] \right\}\ , \label{eq:3.34}
\end{equation}
where $r_0$ can be taken arbitrarily large, so that the expression
(\ref{eq:3.34}) depends only on the asymptotic expansion of $W({\bf x})$
for $|{\bf x}|\to \infty$, say
\begin{equation}
 W ({\bf x}) = \sum_{p\geq 1,\,\ell\geq 0} W^p_L\,
  {\hat n^L\over r^p} \ . \label{eq:3.35}
\end{equation}
Although this is not {\it a priori} evident from its definition
(\ref{eq:3.31}), the explicit expression we shall derive below for $W({\bf
x})$ shows that the multipolar expansion (\ref{eq:3.35}) proceeds according
to the inverse powers of $r$, without involving logarithms. [Actually, the
presence of logarithms in (\ref{eq:3.35}) would not change the result
below.] As was already mentioned above, the only terms in the multipolar
expansion (\ref{eq:3.35}) which can generate a simple pole in $B$ so as to
cancel the factor $B$ in front are those which correspond to an {\it
homogeneous} solution of the Laplace equation $\propto \hat{n}^L /
r^{\ell+1}$. Let us then {\it define} the ``homogeneous'' piece of $W$ by
restricting the double sum in (\ref{eq:3.35}) to the case where $p = \ell +
1$, say (with the definition $W^{\rm hom}_L \equiv W^{\ell+1}_L$)
\begin{equation}
W^{\rm hom}({\bf x}) \equiv \sum_{\ell \geq 0} W^{\rm hom}_L
  {\hat n^L\over r^{\ell+1}} \ . \label{eq:3.36}
\end{equation}
Note that this definition is independent of the choice of the origin ${\bf
x} = 0$ around which one is expanding. This stems from the easily verified
fact that under a constant shift ${\bf x} \to {\bf x} + {\bf c}$ the
homogeneous and inhomogeneous pieces of $W$ do not mix.
Using $\int^{\infty}_{r_0} dr r^{B-1} = - B^{-1} r_0^B = - B^{-1} +
O(B^0)$ and $\int d \Omega\,\hat{n}_L \hat{n}_{L'} \hat{T}_{L'}
= [4 \pi\ell!/(2\ell+1)!!] \hat{T}_L$ (where $n!! \equiv n(n-2)(n-4)$\dots
(1 or 2)) we get the following link between $I^{(W)}_L$ and the
coefficients appearing in the multipolar expansion (\ref{eq:3.36}) of
$W^{\rm hom} ({\bf x})$,
\begin{equation}
GI^{(W)}_L = {\ell !\over (2\ell -1) !!} W^{\rm hom}_L \ . \label{eq:3.37}
\end{equation}
Inserting back (\ref{eq:3.37}) into (\ref{eq:3.36}) shows that the
multipolar expansion of $W^{\rm hom}$ can be compactly written in a form which
is familiar~:
\begin{equation}
W^{\rm hom} = G \sum_{\ell \geq 0} {(-)^{\ell}\over \ell !} I^{(W)}_L
\partial_L {1\over r} \ . \label{eq:3.38}
\end{equation}
The form (\ref{eq:3.38}) is related to another, more convenient, expression
for $I^{(W)}_L$ [in terms of the ``source'' of $W^{\rm hom}$] that we shall
derive below. Let us now obtain an explicit expression for $W ({\bf x})$ in
the point-mass limit and extract its homogeneous part.

Inserting $U = Gm_1/r_1 + Gm_2/r_2$ in the definition of $W_{ij}$ yields
(with $n^i_1 \equiv (x^i-y^i_1)/r_1$, etc.)
\begin{equation}
\triangle W_{ij} = G^2 \left[ m^2_1 {n^i_1 n^j_1\over r^4_1} + m^2_2 {n^i_2
n^j_2\over r_2^4} + m_1 m_2 \left( \partial_{y^i_1} \partial_{y^j_2}
+ \partial_{y^j_1} \partial_{y^i_2} \right) {1\over r_1 r_2}\right] \ ,
\label{eq:3.39}
\end{equation}
whose solution can be written as
\begin{equation}
W_{ij} = G^2 [m^2_1 W^{11}_{ij} + m^2_2 W^{22}_{ij} + m_1 m_2
W^{12}_{ij}] \ , \label{eq:3.40}
\end{equation}
with
\begin{mathletters}
\label{eq:3.41}
\begin{eqnarray}
W^{11}_{ij} &=& {1\over 8} \partial_{ij} \ln r_1 + {1\over 8}
{\delta^{ij}\over r^2_1} \ , \label{eq:3.41a} \\
W^{22}_{ij} &=& {1\over 8} \partial_{ij} \ln r_2 + {1\over 8}
{\delta^{ij}\over r^2_2} \ , \label{eq:3.41b} \\
W^{12}_{ij} &=& {}_ig_j + {}_jg_i  \ . \label{eq:3.41c}
\end{eqnarray}
\end{mathletters}
In Eqs.~(\ref{eq:3.41a}), (\ref{eq:3.41b}) the derivatives $\partial_i
\equiv \partial/\partial x^i$, while in Eq.~(\ref{eq:3.41c}) $g$ is
the quadratic kernel \cite{DI91a}
\begin{mathletters}
\label{eq:3.42}
\begin{eqnarray}
 g( {\bf x}; {\bf y}_1, {\bf y}_2) &\equiv & \ln (r_1 +r_2 +r_{12})\ ,
  \label{eq:3.42a}\\
 \Delta_x g &=& {1\over r_1 r_2}\ , \label{eq:3.42b}
\end{eqnarray}
\end{mathletters}
and we have introduced the same abbreviated notation as above:
\begin{equation}
 {}_ig_j \equiv {\partial\over \partial y^i_1} {\partial\over \partial y^j_2}
  g \ . \label{eq:3.43}
\end{equation}
{}From (\ref{eq:3.40}) we get the cubically nonlinear effective source term
\begin{equation}
 W_{ij} \partial_{ij} U = G^3 [m^3_1 {\cal A}^{111} + m^2_1 m_2
 {\cal A}^{112} + m_1 m_2^2 {\cal A}^{122} + m^3_2 {\cal A}^{222}]
 \ , \label{eq:3.44}
\end{equation}
with
\begin{eqnarray}
 {\cal A}^{111} &=& W_{ij}^{11} \partial_{ij} {1\over r_1}
    = {1\over 8} \partial_{ij} \ln r_1 \partial_{ij} {1\over r_1} =
    - {1\over 2r^5_1}   \ , \label{eq:3.45}\\
 {\cal A}^{112} &=& W_{ij}^{12} \partial_{ij} {1\over r_1} + W^{11}_{ij}
  \partial_{ij} {1\over r_2} = 2\,{}_ig_j \partial_{ij} {1\over r_1}
  +{1\over 8} \partial_{ij} \ln r_1 \partial_{ij} {1\over r_2} + {1\over 8}
  {1\over r^2_1} \Delta {1\over r_2}\ , \label{eq:3.46}
\end{eqnarray}
the other terms in (\ref{eq:3.44}) being obtained by exchanging the roles
of ${\bf y}_1$ and ${\bf y}_2$.

 Note that the two ${\bf x}$-derivatives acting on $1/r_1$ and $1/r_2$
in (\ref{eq:3.46}) [which introduce non locally integrable singularities
in ${\bf x}$-space] are left uneffected and must be interpreted in the sense
of distribution theory (in ${\bf x}$-space). Rigorously speaking, one
is not allowed to work within the framework of distribution theory
(because one is dealing with the product of a distribution
$\partial_{ij}r^{-1}_1$ by a function $_ig_j$ which is not smooth at
${\bf x} ={\bf y}_1$). One should, e.g., use the well-defined analytic
continuation procedure of \cite{D83a} which has been shown to correctly
describe the self-gravity effects of compact bodies. [The latter analytic
continuation procedure yields in particular an unambiguous treatment of
the self source term ${\cal A}^{111}$, Eq.~(\ref{eq:3.45}).] In practice,
a technically easier way to deal with this subtlety is to work with the
${\bf y}_1-$ and ${\bf y}_2-$derivatives of quantities which are less
singular in ${\bf x}$-space (see the first terms in the definitions
of $H$ and $K$ below). When doing so, there arises only one term which
is not well-defined in the sense of distribution theory, and this term,
$\propto {\bf n}_1 \delta ({\bf x}-{\bf y}_1)$, clearly vanishes when
treated more properly by analytic continuation.
Computing $\Delta^{-1} (W_{ij} \partial_{ij} U)$ is easy for ${\cal
A}^{111}$ (using $\Delta r^{-3}_1 =6\, r^{-5}_1$) and the last term
in ${\cal A}^{112}$ (using $r^{-2}_1 \Delta r^{-1}_2 = r^{-2}_{12} \Delta
r^{-1}_2 = \Delta (r^{-2}_{12} r^{-1}_2)$). The other contributions to
${\cal A}^{112}$ are much more intricate to deal with. We succeeded
in evaluating explicitly $\Delta^{-1} {\cal A}^{112}$ by combining the
results of Refs.~\cite{Car65,O73} which pointed out the usefulness of
considering certain ${\bf y}_1-$ and ${\bf y}_2-$derivatives of
combinations of $g$, $\ln r_1$, $r_1$, $r_2$ and $r_{12}$, with the fact
(contained in a somewhat roundabout way in Ref.~\cite{S87}) that the
inverse  Laplacian of $r^{-4}_1 r^{-1}_2$ has a simple expression:
\begin{equation}
 \Delta^{-1} \left( {1\over r^4_1 r_2} \right) = {1\over 2r^2_{12}}
   {r_2\over r^2_1}\ . \label{eq:3.47}
\end{equation}
[The latter being most simply obtained from its easily verified
``inverse": $\Delta (r_2/r^2_1) = 2r^2_{12}/(r^4_1r_2).$]
Denoting $\nabla_1 \equiv \partial/\partial {\bf y}_1$,
$\Delta_1 \equiv \partial/\partial y^i_1 \partial/\partial y^i_1$,
$\nabla_1 \cdotp \nabla_2
\equiv \partial/\partial y^i_1 \partial/\partial y^i_2$, and $n_{12}^i
\equiv (y^i_1 -y^i_2)/r_{12}$, we define the quantities
\begin{eqnarray}
H &\equiv &\Delta_1 (\nabla_1\cdotp\nabla_2) \left[ {r_1+r_{12}\over 2}
  \, g \right] - {1\over 2} {r_2\over r^2_{12} r^2_1} \nonumber\\
 && + {1\over 2} {1\over r_{12}r_1^2} + {{\bf n}_{12}\cdotp {\bf n}_1\over
   r_{12} r^2_1} + {1\over r^2_{12} r_1}\nonumber \\
 && + {3\over 2} {{\bf n}_{12} \cdotp {\bf n}_1\over r^2_{12}r_1}
  - {{\bf n}_{12}\over r^2_{12}} \cdotp \nabla_1\, g\ ,
     \label{eq:3.48}\\
 K &\equiv & (\nabla_1\cdotp \nabla_2)
  \left[ {1\over r_2} \ln r_1  - {1\over r_2}\ln r_{12} \right] \nonumber \\
 && +{1\over 2} {r_2\over r^2_{12} r^2_1} - {1\over 2} {1\over r^2_1 r_2} +
   {1\over 2} {1\over r^2_{12} r_2}\ , \label{eq:3.49}
\end{eqnarray}
which verify
\begin{eqnarray}
\Delta_x H &=& 2\,{}_ig_j\,\partial_{ij} {1\over r_1}\ , \label{eq:3.50}\\
\Delta_x K &=& 2\,\partial_{ij} \ln r_1\,\partial_{ij} {1\over r_2} \ .
   \label{eq:3.51}
\end{eqnarray}
Finally, introducing the combination
\begin{equation}
  Q \equiv 4H + {1\over 4} K + {1\over 2} {1\over r^2_{12} r_2}\ ,
  \label{eq:3.52}
\end{equation}
we find that $W({\bf x})$, the unique solution of (\ref{eq:3.31a}) falling
off at infinity, is given by
\begin{equation}
 W = {G^3\over c^4} \left[ {m^3_1\over 3r^3_1} - m^2_1m_2Q
  + (\hbox{two other terms obtained by exchanging }1\leftrightarrow 2)
       \right] \ .  \label{eq:3.53}
\end{equation}

To project out from $W({\bf x})$ the part $W^{\rm hom}$ whose multipolar
expansion, when $|{\bf x}|\to \infty$, is purely homogeneous, it is
convenient, besides using the defining criterion that it contains only
terms of the type $\hat n^L/|{\bf x}|^{\ell+1}$, to notice that $W^{\rm hom}$
must also be a non smooth function of ${\bf y}_1$ and ${\bf y}_2$
(considered jointly). This can be shown either from the general
structure of our generation formalism, or, more simply, by remarking,
on dimensional and tensorial grounds, that $I^{(W)}_L$ must be of the
form: $G^2m^p_1 m^{3-p}_2 /(c^4r^2_{12})$ times a tensor product of
$\ell$ vectors ${\bf y}_1$ or ${\bf y}_2$. When using these two criteria
in conjunction, one finds that many terms in $H$ and $K$ project out to
zero. For instance, by explicating
\begin{equation}
 \Delta_1 \left[ {r_1 +r_{12}\over 2} g\right] = \left( {1\over r_1}
 + {1\over r_{12}} \right) g + {1\over r_1} + {1\over r_{12}} -
 {r_2\over r_{12}r_1} \label{eq:3.54}
\end{equation}
with the decomposition $g=k+{1\over 2} h$ (see Ref.~\cite{DI91a}) where
$k$ is smooth in $({\bf y}_1, {\bf y}_2)$ and where
\begin{mathletters}
\label{eq:3.55}
\begin{eqnarray}
 h ({\bf x};{\bf y}_1, {\bf y}_2) &\equiv & \ln \left( {r_1+r_2+r_{12}
  \over r_1+r_2-r_{12}} \right)\ , \label{eq:3.55a}\\
 \Delta_x h &=& - 4\pi \delta_{12}\ , \label{eq:3.55b}
\end{eqnarray}
\end{mathletters}
has a purely homogeneous multipolar expansion, one finds that the first
term in $H$ (Eq.~(\ref{eq:3.48})) projects out to zero.
 Finally one gets
\begin{eqnarray}
 W^{\rm hom} &=& {G^3m^2_1m_2 \over c^4} \left[ {1\over 4}
  \nabla_1 \cdotp \nabla_2
\left( {\ln r_{12}\over r_2} \right) \right. \nonumber \\
 && + {15\over 8r^2_{12}} \left( {r_2\over r^2_1}\right)^{\rm hom}
 - 4\, {{\bf n}_{12}\over r_{12}} \cdotp {{\bf n}_1\over r_1^2}
 - {4\over r^2_{12} r_1} \nonumber \\
 && \left.-{5\over 8}{1\over r^2_{12} r_2} + 2\,{{\bf n}_{12}\over r^2_{12}}
  \cdotp \nabla_1 h\right] + (1\leftrightarrow 2)\ ,
\label{eq:3.56}
\end{eqnarray}
where $(r_2/r^2_1)^{\rm hom}$ denotes the homogeneous projection of
$r_2/r^2_1$ and where $h$ was defined in Eqs.~(\ref{eq:3.55}). Let us
now evaluate $u\equiv (r_2/r^2_1)^{\rm hom}$. By expanding $r_2 =|{\bf
x}-{\bf y}_2| =|{\bf r}_1+{\bf r}_{12}|$ [with ${\bf r}_1 ={\bf x}
-{\bf y}_1$, ${\bf r}_{12} ={\bf y}_1-{\bf y}_2$] in powers of
${\bf r}_{12}$, we get the expansion at infinity of $r_2/r^2_1$:
\begin{equation}
 {r_2\over r^2_1} = \sum_{\ell \geq 0} {1\over \ell!} r^L_{12}
  {\partial_Lr_1\over r^2_1} \label{eq:3.57}
\end{equation}
[where, as usual, $r^L_{12}\equiv r^{i_1}_{12}\dots r^{i_\ell}_{12}$].
The projection of
$\partial_Lr_1/r^2_1$ onto the homogeneous solutions $\propto \hat n^L_1
/r^{\ell+1}_1$ (using ${\bf x} ={\bf y}_1$ as origin for the expansion
at infinity) is simply obtained by taking the STF projection of the
multi-index $L$. Hence (with $(-1)!!=1$, $(-3)!!=-1$)
\begin{equation}
 u \equiv \left( {r_2\over r^2_1}\right)^{\rm hom} = \sum_{\ell\geq 0}
 {1\over \ell!} r^L_{12} {\hat\partial_Lr_1\over r^2_1} = \sum_{\ell\geq 0}
{(-)^{\ell-1}\over\ell!} (2\ell-3)!!\,r^L_{12} {\hat n^L_1\over r_1^{\ell+1}}
 \ . \label{eq:3.58}
\end{equation}
Using formula (A25) from \cite{BD86} this can be rewritten in terms of
Legendre polynomials of ${\bf n}_1\cdotp {\bf n}_{12}$:
\begin{equation}
 u = {1\over r_{12}} \sum_{\ell\geq 0} {(-)^{\ell-1}\over 2\ell-1}
 P_\ell ({\bf n}_1 \cdotp {\bf n}_{12}) \left( {r_{12}\over r_1}\right)
 ^{\ell+1}\ . \label{eq:3.59}
\end{equation}
It is clear from Eqs.~(\ref{eq:3.58})-(\ref{eq:3.59}) that $u({\bf x},
{\bf y}_1, {\bf y}_2)$ is a solution of $\Delta_x u=0$ which has an
axial symmetry around the straight line joining ${\bf y}_1$ to ${\bf
y}_2$.  We can represent $u$ in closed form by introducing a
distribution of ``charges" along the segment ${\bf y}_1 -{\bf y}_2$, say
\begin{equation}
 u({\bf x}; {\bf y}_1, {\bf y}_2) = \int^1_0 d\alpha\,
  {w(\alpha) \over |{\bf x} -{\bf y}_\alpha|}\ , \qquad
 {\bf y}_\alpha\equiv (1-\alpha) {\bf y}_1 +\alpha {\bf y}_2\ .\label{eq:3.60}
\end{equation}
By identifying (\ref{eq:3.60}) and (\ref{eq:3.59}) on the axis of symmetry,
we see that the weight $w(\alpha)$ with which the ``charges'' are
distributed on the segment ${\bf y}_1 -{\bf y}_2$ must satisfy
\begin{equation}
 \int^1_0 d\alpha\, w(\alpha) \alpha^\ell = - {1\over 2\ell-1}\ .
  \label{eq:3.61}
\end{equation}
We find that $w(\alpha)$ must be defined within the framework of
distribution theory (rather than that of ordinary, locally integrable
functions) as
\begin{equation}
 w(\alpha) = Pf \left[ -{1\over 2} \alpha^{-3/2} \right]\ ,\label{eq:3.62}
\end{equation}
where the symbol $Pf$ denotes Hadamard's partie finie for an integral
over the interval $0\leq \alpha \leq 1$ (see Ref.~\cite{Schwartz}).
Actually, this could also be written in terms of the
analytic-continuation Finite Part operator used in our general
formalism, but Hadamard's $Pf$ operator is a simpler object when, as is
the case in Eq.~(\ref{eq:3.62}), the integrals to be defined involve
non-integer powers. We want also to emphasize by the change of notation
that the appearance of the distribution $w(\alpha)$ is quite
disconnected from the formalism behind our starting formula
(\ref{eq:2.17}). We note in passing that we could dispense
of using Hadamard's partie finie by using $w(\alpha) =(d/d\alpha)
(\alpha^{-1/2})$, where the derivative is taken in the sense of
distributions, and by integrating (\ref{eq:3.60}) by parts. However, the
form (\ref{eq:3.62}) is technically more convenient for our purpose.

 Summarizing the results so far, we have succeeded in resumming the
infinite multipolar series defining $W^{\rm hom}$, Eq.~(\ref{eq:3.36}),
to obtain it as a {\it finite} sum of derivatives of $r^{-1}_1$ and
$r^{-1}_2$ [see Eq.~(\ref{eq:3.56})], plus two more complicated
homogeneous solutions: a ${\bf y}_1$-gradient of the function $h$,
Eqs.~(\ref{eq:3.55}), and the function $u$, which is a distributional
superposition of elementary solutions $|{\bf x}-{\bf y}_\alpha|^{-1}$
on the segment ${\bf y}_1 - {\bf y}_2$. Note that while Eq.~(\ref{eq:3.36})
defined only the asymptotic expansion at infinity of $W^{\rm hom}$, our
closed-form result for $W^{\rm hom} ({\bf x})$ defines it for all values
of ${\bf x}$.  At this stage, it is convenient to bypass the link
(\ref{eq:3.37})-(\ref{eq:3.38}) between the object we seek, $I^{(W)}_L$,
and the coefficients of the multipolar expansion of $W^{\rm hom} ({\bf x})$
by using our detailed results on $W^{\rm hom}$ for defining the
distributional ``source'' (in {\bf x}-space) of $W^{\rm hom}$ by
\begin{equation}
 S ({\bf x}) \equiv - {1\over 4\pi} \Delta_x W^{\rm hom} ({\bf x})\ .
  \label{eq:3.63}
\end{equation}
Thanks to our resummation of $W^{\rm hom}$ as a sum of elementary
``homogeneous'' (in the sense of {\it functions}, but not of {\it
distributions}) solutions, the definition (\ref{eq:3.63}) makes sense
with $S({\bf x})$ being a distribution (in {\bf x}-space) whose support
is localized on the segment joining ${\bf y}_1$ to ${\bf y}_2$. More
precisely, we have
\begin{eqnarray}
  S &=& {G^3 m^2_1m_2\over c^4} \left[ {1\over 4} \nabla_1 \cdotp
  \nabla_2 (\ln r_{12} \delta ({\bf x} -{\bf y}_2)) \right. \nonumber \\
 && -{15\over 16} {1\over r^2_{12}} Pf \int^1_0 d\alpha\, \alpha^{-3/2}
     \delta ({\bf x}-{\bf y}_\alpha)
   - 4 {{\bf n}_{12}\over r_{12}} \cdotp \nabla_1 \delta
     ({\bf x}-{\bf y}_1) \nonumber \\
 && - {4\over r^2_{12}} \delta ({\bf x}-{\bf y}_1)
    - {5\over 8} {1\over r^2_{12}} \delta ({\bf x}-{\bf y}_2)\nonumber \\
&& \left.  + 2 {{\bf n}_{12}\over r^2_{12}} \cdotp
   \nabla_1 \left( r_{12} \int^1_0 d\alpha\, \delta
   ({\bf x}-{\bf y}_\alpha)\right) \right]
 + (1 \leftrightarrow 2)\ . \label{eq:3.64}
\end{eqnarray}
 The introduction of the source $S$ of $W^{\rm hom}$ simplifies our
evaluation of the multipole moments $I^{(W)}_L$. Indeed, from the definition
(\ref{eq:3.36}) and the fact that $S({\bf x})$ has a {\it compact} support
we deduce as usual
\begin{equation}
 W^{\rm hom} ({\bf x}) = \int d^3{\bf y} {S({\bf y})\over |{\bf x}-{\bf y}|}
 = \sum_{\ell \geq 0} {(-)^\ell\over \ell!}\partial_L {1\over |{\bf x}|}
  \int d^3{\bf y}\, y^L  S({\bf y}) \ . \label{eq:3.65}
\end{equation}
By comparing with (\ref{eq:3.38}), we find
\begin{equation}
 GI^{(W)}_L = \int d^3 {\bf x}\, \hat x^L S({\bf x})\ . \label{eq:3.66}
\end{equation}
It is interesting to remark that the work of the present subsection,
together with the one of the previous subsection, is analogous to the
method used in Ref.~\cite{DI91a}. Basically, we have decomposed
$-4W_{ij}\partial_{ij}U/c^4$, which was an effective source term for
the inner metric (i.e. a cubically nonlinear analog of the quadratically
nonlinear terms obtained from the elementary object $r^{-1}_1 r^{-1}_2$)
into a piece $\Delta_x W^{\rm hom} = -4\pi S$ which has a compact
support (analog to $\Delta_x h=-4\pi \delta_{12}$, in \cite{DI91a},
defining the compact source $\tau_c^{\mu\nu}$) and a complementary piece
$\Delta_x W^{\rm inhom}$, with $W^{\rm inhom} \equiv W-W^{\rm hom}$
being the analog of the kernel $k$, which does not contribute to
(\ref{eq:2.17}).  In other words, the combination of the results of
Ref.~\cite{B94} with the results given here, end up by yielding a
formula for $I_L$ given {\it explicitly} by an integral over a {\it
compact support} (after the two replacements $r^{-1}_1r^{-1}_2 \to -2\pi
\delta_{12}$ and $W_{ij} \partial_{ij} U/c^4\to \pi S$ in
Eq.~(\ref{eq:2.17})).  From Eq.~(\ref{eq:3.66}) one can read off
directly from (\ref{eq:3.64}) an explicit expression for $I^{(W)}_L$.
Note that one can get rid of the logarithm appearing in the first term
of this expression by explicating the $\nabla_1$ derivative:
\begin{equation}
{1\over 4} (\nabla_1\cdotp \nabla_2) (\ln r_{12}\,\hat y_2^L)
 = - {1\over 4r^2_{12}} \hat y^L_2 + {1\over 4} {{\bf n}_{12}\over r_{12}}
 \cdotp \nabla_2 \hat y^L_2 \ . \label{eq:3.67}
\end{equation}
This leads to our (essentially) final result
\begin{eqnarray}
 I^{(W)}_L &=& {G^2m^2_1m_2\over c^4} \left[ - {4\over r^2_{12}} \hat y^L_1
 - {7\over 8r^2_{12}} \hat y^L_2 \right. \nonumber \\
 && - 4 {{\bf n}_{12}\over r_{12}} \cdotp \nabla_1 \hat y^L_1 + {1\over 4}
 {{\bf n}_{12}\over r_{12}} \cdotp \nabla_2 \hat y^L_2\nonumber\\
 && + 2 {{\bf n}_{12}\over r^2_{12}} \cdotp \nabla_1 \left( r_{12} \int^1_0
 d\alpha \hat y^L_\alpha \right)\nonumber \\
 && \left. - {15\over 16} {1\over r^2_{12}} Pf \int^1_0 d\alpha\,
 \alpha^{-3/2} \hat y^L_\alpha \right] + (1\leftrightarrow 2)\ ,
 \label{eq:3.68}
\end{eqnarray}
where, as usual, $\hat y^L_\alpha \equiv y_\alpha^{<i_1} \cdots y^{i_\ell>}
_\alpha$, and where, because of the dissymmetry betweeen ${\bf y}_1$ and
${\bf y}_2$ in the last integral, it is important to recall the
definition used here: ${\bf y}_\alpha \equiv (1-\alpha) {\bf y}_1 +
\alpha {\bf y}_2$ (which differs from the notation used in \cite{DI91a}
by the replacement $\alpha\to 1-\alpha$).

 To get a completely explicit expression for $I^{(W)}_L$ in terms of
${\bf y}_1$ and ${\bf y}_2$ one needs first to compute the integrals
appearing in (\ref{eq:3.68}): the first one has been given in
Eq.~(\ref{eq:3.23}) above (actually, one uses here the equivalent
expression given by Eq.~(5.24) in \cite{DI91a}), the second one is simply
obtained by expanding the tensorial power $y^L_\alpha = ({\bf y}_1
-\alpha {\bf y}_{12})^{\otimes L}$ in powers of $\alpha$ and by using
the elementary integrals (\ref{eq:3.61}).  Then the derivatives with
respect to ${\bf y}_1$ and ${\bf y}_2$ must be effected.  One can
drastically cut down the whole calculation by using the fact that
$I^{(W)}_L$ must (as is clear from (\ref{eq:3.66})) transform under
global spatial shifts of the coordinate system, ${\bf x}\to {\bf x}+
\mbox{\boldmath$\varepsilon$}$, ${\bf y}_A \to {\bf y}_A +
\mbox{\boldmath$\varepsilon$}$ according to
$\delta_{\mbox{\boldmath$\varepsilon$}} I^{(W)}_L = \ell
\varepsilon_{<i_\ell} I^{(W)}_{L-1>} + O(\varepsilon^2)$ (see also
Appendix B of \cite{DI91a}).  Defining for brevity some $\ell$-dependent
coefficients,
\begin{mathletters}
\label{eq:3.69}
\begin{equation}
 C_\ell \equiv 2\delta^0_\ell + 4 \delta^1_\ell + (-)^\ell
  {\ell^2+3\ell +2\over 2(2\ell-1)}\  \label{eq:3.69a}
\end{equation}
(where $\delta^0_\ell =1$ if $\ell=0$ and 0 otherwise, and similarly
for $\delta^1_\ell$), we obtain finally
\begin{equation}
 I^{(W)}_L = - {G^2m^2_1m_2\over c^4r^2_{12}} \sum^\ell_{p=0}
 {\ell \choose p} C_p \ y^{<L-P}_1 y^{P>}_{12} + (1\leftrightarrow 2)
  \ , \qquad {\ell\choose p} \equiv {\ell!\over p!(\ell -p)!}\ ,
\label{eq:3.69b}
\end{equation}
\end{mathletters}
where the multi-indices are such that there are $\ell -p$ indices
on ${\bf y}_1$ and $p$ on ${\bf y}_{12} ={\bf y}_1 -{\bf y}_2$.
Writing out explicitly the low orders in $\ell$ we find
\begin{equation}
 I^{(W)}_L = - {G^2m^2_1m_2\over c^4r^2_{12}} \
 \overline Q_L + (1\leftrightarrow 2)   \ , \label{eq:3.70}
\end{equation}
with
\begin{mathletters}
\label{eq:3.71}
\begin{eqnarray}
  \overline Q &=& 1\ , \label{eq:3.71a} \\
  \overline Q_i &=& y^i_1 + y^i_{12}\ , \label{eq:3.71b} \\
  \overline Q_{ij} &=& y^{<ij>}_1 + 2y^{<i}_1 y^{j>}_{12}
   + 2 y^{<ij>}_{12}\ , \label{eq:3.71c} \\
  \overline Q_{ijk} &=& y^{<ijk>}_1 + 3y^{<ij}_1 y^{k>}_{12}
   + 6y^{<i}_1 y^{jk>}_{12} - 2 y^{<ijk>}_{12}\ , \label{eq:3.71d} \\
  \overline Q_{ijkl} &=& y^{<ijkl>}_1 + 4y^{<ijk}_1 y^{l>}_{12}
   +12y^{<ij}_1 y^{kl>}_{12} - 8 y^{<i}_1 y^{jkl>}_{12}
  + {15\over 7} y^{<ijkl>}_{12}\ . \label{eq:3.71e}
\end{eqnarray}
\end{mathletters}
One should note that in all the formulas for $W$ and $I^{(W)}_L$ given
above, starting with Eq.~(\ref{eq:3.53}), one must complement the
explicitly written results by adding similar terms, proportional to
$m_1 m^2_2$, obtained by exchanging the labels 1 and 2 (remember that
$y^i_{21} \equiv y^i_2 -y^i_1 =-y^i_{12}$). For instance, the complete
expression of the quadrupole reads
\begin{equation}
 I^{(W)}_{ij} = -{G^2m_1m_2\over c^4r^2_{12}}\, {\rm STF}_{ij}\,
 [ m_1y^{ij}_1 + m_2y^{ij}_2 +2 (m_1y^i_1 -m_2y^i_2) y^j_{12}
  +2 (m_1+m_2) y^{ij}_{12} ]\ . \label{eq:3.72}
\end{equation}
An independent, direct derivation of the quadrupole (\ref{eq:3.72})
(without using the decomposition of $W$ into homogeneous and inhomogeneous
parts) is given in
Appendix~\ref{sec:apb}. In the center of mass frame, we finally get for
the cubically nonlinear contribution to the quadrupole
\begin{equation}
 I^{(W)}_{ij} = -\nu m \gamma^2 (2+5\nu) \hat x^{ij}\ . \label{eq:3.73}
\end{equation}
Summing up the explicit results obtained for the three pieces of the
2PN-accurate quadrupole (considered for circular orbits, in the
center-of-mass frame) we get
\begin{eqnarray}
I_{ij}= I^{(C)}_{ij} + I^{(Y)}_{ij} + I^{(W)}_{ij}
&=&{\rm STF}_{ij}\,
\nu m\left\{ x^{ij}\;-\;\frac{\gamma}{42}(1+39\nu)x^{ij}\;+\;
\frac{11}{21}\frac{r^2}{c^2}v^{ij}(1-3\nu)\;\right.\nonumber\\
&&\;-\;\;\frac{\gamma^{2}x^{ij}}{1512}(461+18395\nu+241\nu^2)\nonumber\\
&&\left.\;+\;\frac{\gamma r^{2}v^{ij}}{378c^2}(1607-1681\nu+229\nu^2)\right\}
\ .\label{eq:3.74}
\end{eqnarray}

\section{The 2PN-accurate waveform and energy loss}
\label{sec:4}
\subsection{The waveform including its tail contribution}
\label{sec:4.1}

 The 2PN-accurate waveform is given by Eq.~(\ref{eq:2.3}) in terms of
the ``radiative'' multipole moments $U_L$ and $V_L$ which are in turn
linked to the source moments $I_L$ and $J_L$ by Eqs.~(\ref{eq:2.6}) and
(\ref{eq:2.7}). The latter equations involve some tail integrals and
therefore yield a natural decomposition of the waveform into two pieces,
one which depends on the state of the binary at the retarded instant
$T_R\equiv T-R/c$ only (we qualify this piece as ``instantaneous''), and
one which is {\it a priori} sensitive to the binary's dynamics at all
previous instants $T_R-\tau \leq T_R$ (we refer to this piece as the
``tail'' contribution).  More precisely, we decompose
\begin{equation}
  h^{TT}_{km} = (h^{TT}_{km})_{\rm inst}
       + (h^{TT}_{km})_{\rm tail}\ , \label{eq:4.1}
\end{equation}
where the ``instantaneous'' contribution is defined by
\begin{eqnarray}
 (h^{TT}_{km})_{\rm inst} = {2G\over c^4R} {\cal P}_{ijkm}
     \biggl\{ I^{(2)}_{ij}
  &+& {1\over c} \left[ {1\over 3} N_a I^{(3)}_{ija} + {4\over 3}
   \varepsilon_{ab(i} J^{(2)}_{j)a} N_b \right] \nonumber \\
  &+& {1\over c^2} \left[ {1\over 12} N_{ab} I^{(4)}_{ijab} + {1\over 2}
   \varepsilon_{ab(i} J^{(3)}_{j)ac} N_{bc} \right] \nonumber \\
  &+& {1\over c^3} \left[ {1\over 60} N_{abc} I^{(5)}_{ijabc} + {2\over 15}
   \varepsilon_{ab(i} J^{(4)}_{j)acd} N_{bcd} \right] \nonumber \\
  &+& {1\over c^4} \left[ {1\over 360} N_{abcd} I^{(6)}_{ijabcd} + {1\over 36}
   \varepsilon_{ab(i} J^{(5)}_{j)acde} N_{bcde} \right] \biggr\}
   \ ,\label{eq:4.2}
\end{eqnarray}
and where the ``tail'' contribution reads
\begin{mathletters}
\label{eq:4.3}
\begin{eqnarray}
 (h^{TT}_{km})_{\rm tail} = {2G\over c^4R} {2Gm\over c^3}{\cal P}_{ijkm}
\int^{+\infty}_0 &d\tau& \left\{ \ln \left( {\tau\over 2b_1}\right)
    I^{(4)}_{ij} (T_R -\tau)\right. \nonumber\\
 && \quad + {1\over 3c} \ln \left( {\tau\over 2b_2}\right)
  N_a I^{(5)}_{ija} (T_R-\tau) \nonumber\\
 && \quad \left. + {4\over 3c} \ln \left( {\tau\over 2b_3}\right)
 \varepsilon_{ab(i} N_b J^{(4)}_{j)a} (T_R -\tau) \right\} \ .
 \label{eq:4.3a}
\end{eqnarray}
 We have used for simplicity the notation
\begin{equation}
 b_1 \equiv b\, e^{-11/12}\ , \qquad  b_2 \equiv b\, e^{-97/60}\ , \qquad
 b_3 \equiv b\, e^{-7/6}\ . \label{eq:4.3b}
\end{equation}
\end{mathletters}
Note that the decomposition (\ref{eq:4.1})-(\ref{eq:4.3}) into instantaneous
and tail contributions is convenient but by no means unique.  In
particular the logarithms of the constants $b_1$, $b_2$, $b_3$ in
Eq.~(\ref{eq:4.3a}) could be as well transferred to the instantaneous
side (\ref{eq:4.2}).

We first study the instantaneous contribution (\ref{eq:4.2}), which is
straightforwardly computed from the 2PN-accurate mass quadrupole moment
previously derived (Eq.~(\ref{eq:3.74})), and from the knowledge of the
other moments necessitating the 1PN accuracy at most.  The 1PN-accurate
source moments have been derived in \cite{BD89} (see Eq.~(\ref{eq:2.9})
above) and \cite{DI91a}.  Note that the result (5.18) of \cite{DI91a} is
equivalent to inserting Eq.~(\ref{eq:3.21}) into the result (\ref{eq:2.13})
above. We list below the relevant mass-type and
current-type moments which have to be inserted into the waveform
(\ref{eq:4.2}):
\begin{mathletters}
\label{eq:4.4}
\begin{eqnarray}
I_{ij} &=& \nu m\, {\rm STF}_{ij}
\left\{ x^{ij} - \frac{\gamma}{42}\;(1+39\nu)\;x^{ij} + \frac{11}{21}\;
\frac{r^2}{c^2}v^{ij}\;(1-3\nu)\right.\nonumber   \\
&&- \frac{\gamma^2 x^{ij}}{1512}\;\;\left(461+18395\nu+241\nu^2\right)
\nonumber  \\
&&+\left. \frac{\gamma r^2 v^{ij}}{378c^2}\;\;\left(1607-1681\nu+229\nu^2
\right) \right\} \ ,\label{eq:4.4a}\\
I_{ijk} &=& \nu m\;\;( X_2 - X_1 )\, {\rm STF}_{ijk}
\left\{ x^{ijk} - \gamma\nu x^{ijk}
+ \frac{r^2}{c^2}\;\;v^{ij} x^k ( 1-2\nu ) \right\}\ ,\label{eq:4.4b}\\
I_{ijkl} &=& \nu m\, {\rm STF}_{ijkl} \left\{ x^{ijkl}
( 1 - 3\nu ) + \frac{\gamma }{110}x^{ijkl}\;\;(3 - 125\nu + 345\nu^2)
 \right.\nonumber \\
&&+\left.\frac{78}{55}\;\;\frac{r^2}{c^2}\;\;v^{ij} x^{kl} ( 1 - 5\nu
     + 5\nu^2 ) \right\}\ , \label{eq:4.4c}\\
I_{ijklm}&=&\nu m(1-2\nu)(X_2-X_1)\,\hat x^{ijklm}\ ,\label{eq:4.4d} \\
I_{ijklmn}&=& \nu m (1-5\nu + 5\nu^2)\,
\hat x^{ijklmn}\ ,\label{eq:4.4e}\\
J_{ij} &=&\nu m(X_2-X_1)\,{\rm STF}_{ij}\,\varepsilon_{jab}\left\{x^{ai}v^b
+ \frac{\gamma}{28}\;\;x^{ai} v^b(67 - 8\nu) \right\}\ ,\label{eq:4.4f}\\
J_{ijk} &=& \nu m\,{\rm STF}_{ijk}\,\varepsilon_{kab} \left\{x^{aij} v^b (1
  - 3\nu ) + \frac{\gamma x^{aij} v^b}{90} (181 - 545\nu + 65\nu^2)
 \right.\nonumber \\
&&+\left. \frac{7}{45}\,\frac{r^2}{c^2}\;\;x^a v^{bij}\;
(1 - 5\nu + 5\nu^2) \right\}\ ,  \label{eq:4.4g}\\
J_{ijkl}&=& \nu m (1-2\nu) (X_2 - X_1)\, {\rm STF}_{ijkl}
\,\varepsilon_{lab}\, x^{aijk} v^b\ ,  \label{eq:4.4h} \\
J_{ijklm}&=& \nu m (1-5\nu + 5\nu^2)\,{\rm STF}_{ijklm}
\,  \varepsilon_{mab}\, x^{aijkl} v^b\ . \label{eq:4.4i}
\end{eqnarray}
\end{mathletters}
Note that Ref.~\cite{DI91a} gave two different forms of the current-type
moments: a ``central'' form, its Eq.~(5.18), equivalent to the form (2.13)
above (taken from \cite{B94}), and a ``potential'' form. We found the
first form simpler to evaluate when one is interested in computing the
moments in the center-of-mass frame. [By using it, we detected an error in
\cite{DI94} which computed the potential form of the octupole $J_{ijk}$:
the coefficient of the twelfth term in Eq.~(9) of Ref.~\cite{DI94}
should read $-5/2$ instead of $+3/2$ (this twelfth term was the only one
not checked by the transformation law of multipole moments under a
constant shift of the spatial origin of the coordinates).] The relevant
time derivatives of the moments (\ref{eq:4.4}), as well as all their
contractions with the unit direction ${\bf N}$ which are needed in the
computation of $(h^{TT}_{km})_{\rm inst}$, are relegated to
Appendix~\ref{sec:apc}. One gets for the 2PN-accurate instantaneous part
of the waveform (in the case of two bodies moving on a circular orbit):
\begin{mathletters}
\label{eq:4.5}
\begin{equation}
(h^{TT}_{km})_{\rm inst} = \frac{2G\nu m}{c^4R}\, {\cal P}_{ijkm}
 \left\{\xi^{(0)}_{ij} + (X_2-X_1) \xi^{(1/2)}_{ij} + \xi^{(1)}_{ij}
+ (X_2-X_1)\, \xi^{(3/2)}_{ij} +
  \xi^{(2)}_{ij}\right\}\ , \label{eq:4.5a}
\end{equation}
where the various post-Newtonian pieces are given by
\begin{eqnarray}
\xi^{(0)}_{ij}&=& 2 (v^{ij} - \frac{Gm}{r} n^{ij})\ ,\label{eq:4.5b}\\
\xi^{(1/2)}_{ij} &=& - 6\frac{Gm}{r} (Nn)
\frac{n^{(i}v^{j)}}{c} - \frac{(v N)}{c}\left \{ \frac{Gm}{r} n^{ij} -
2v^{ij}\right\} \ , \label{eq:4.5c}\\
\xi^{(1)}_{ij} &=& \frac{1}{3}(1-3\nu)
\left[(Nn)^2 \gamma \left\{ 10 \frac{Gm}{r} n^{ij} - 14v^{ij}\right\}
\right. \nonumber \\
&&- 32\left. (Nn)(Nv) \gamma n^{(i} v^{j)}+\frac{(Nv)^2}{c^2}
 \left\{6v^{ij} - 2\frac{Gm}{r} n^{ij}\right\}\right]\nonumber \\
&&- \gamma v^{ij} \left({1\over 3}+\nu\right) + \gamma \frac{Gm}{r}
  n^{ij} \left({19\over 3} - \nu\right)\ , \label{eq:4.5d} \\
\xi^{(3/2)}_{ij} &=& (1-2\nu) \left\{
\frac{65}{6} (Nn)^3 \gamma \frac{Gm}{r} \frac{n^{(i}v^{j)}}{c}\right.
- \frac{46}{3} (Nn) (Nv)^2 \frac{\gamma}{c}  n^{(i}v^{j)}\nonumber \\
&&+ \gamma (Nn)^2 \frac{(Nv)}{c}\left [- \frac{43}{3} v^{ij} + \frac{37}{4}
\frac{Gm}{r} n^{ij}\right]\nonumber \\
&&+\left. \frac{(Nv)^3}{c^3}\left [- \frac{1}{3} \frac{Gm}{r} n^{ij} +
 2v^{ij}\right]\right\}
+ (Nn)\gamma (\frac{95-18\nu}{6}) \frac{Gm}{r} \frac{n^{(i}v^{j)}}{c}\nonumber
\\
&&+\frac{(Nv)}{c}\left [- \frac{2}{3} (1+\nu) \gamma v^{ij} +
\frac{81-2\nu}{12} \gamma \frac{Gm}{r} n^{ij}\right]\ ,\label{eq:4.5e} \\
\xi^{(2)}_{ij} &=&
\gamma^2n^{ij} \left[- \frac{Gm}{r} (\frac{361 +65\nu +
 45\nu^2 }{60} )\right.
 + (Nv)^2 ( \frac{101- 295\nu - 15\nu^2 }{15} )\nonumber \\
&& - \frac{Gm}{r} (Nn)^2 (\frac{309 - 995\nu + 195\nu^2}{15})
+\frac{86}{5} (Nn)^2 (Nv)^2  (1 - 5\nu + 5\nu^2)\nonumber \\
&&-\left.\frac{94}{15}\frac{Gm}{r}(Nn)^4 (1-5\nu +5\nu^2) \right]\nonumber \\
&&+ v^{ij} \left[ -\gamma^2 (  \frac{419 + 1325\nu + 15\nu^2}{60})
- \gamma \frac{(Nv)^2}{c^2} ( 1 - 3\nu - \nu^2 )\right.\nonumber \\
&&+ \gamma^2 (Nn)^2 ( \frac{163 - 545\nu + 135\nu^2}{15} )
+2 \frac{(Nv)^4}{c^4} (1 - 5\nu + 5\nu^2) \nonumber \\
&&+ \frac{128}{15} \gamma^2(Nn)^4
 (1 - 5\nu + 5\nu^2)
-\left.30 \gamma \frac{(Nv)^2(Nn)^2}{c^2}(1-5\nu+5\nu^2) \right]\nonumber \\
&&+ n^{(i}v^{j)} \gamma \left[ \gamma (Nn)(Nv)
   (\frac{176 - 560\nu + 80\nu^2}{5} )
-20 (nN) \frac{(Nv)^3}{c^2}  (1 - 5\nu + 5\nu^2)\right. \nonumber \\
&&+\left.\frac{228}{5}\gamma(Nn)^3 (Nv)(1-5\nu+5\nu^2)\right]\ .\label{eq:4.5f}
\end{eqnarray}
\end{mathletters}
Here ${\bf n}$ is a shorthand for ${\bf x}/r$. Usual euclidean
scalar products are denoted e.g. by $(Nn)={\bf N}\cdotp {\bf n}$. We
recall that the parameter  $\gamma =Gm/(c^2r)$ is related to the relative
velocity ${\bf v}$ of the bodies by Eq.~(\ref{eq:3.13}). When comparing
the waveform (\ref{eq:4.5}) with the less accurate 1.5PN waveform
obtained in Ref.~\cite{W94}, we find several discrepancies between the
coefficients entering the 1.5PN piece $\xi^{(3/2)}_{ij}$
(Eq.~(\ref{eq:4.5e})). We have checked directly, by differentiating
and squaring the complete 2PN waveform (\ref{eq:4.5}) for circular orbits,
that the correct instantaneous part of the energy loss is recovered (i.e.
Eq.~(\ref{eq:4.12}) below). Therefore we suspect that the contentious
coefficients in Ref.\cite{W94} are erroneous.

The tail contribution (4.3) is more difficult to evaluate than the
instantaneous one because it involves an integral extending over the whole
past evolution (or past ``history'') of the binary. In another sense,
however, it is easier to evaluate because it necessitates at the 2PN-level
the knowledge of the quadrupole and octupole moments of the binary with
{\it Newtonian} accuracy only. This is clear owing to the explicit powers
of $1/c$ appearing in Eq.~(\ref{eq:4.3a}). (Only at the 2.5PN level shall
we need to include a post-Newtonian correction into the tail contribution.)
In order to compute the tail integrals, we shall follow a procedure
which is {\it a priori} dangerously formal (although most natural), but
has been fully justified in Ref.~\cite{BS93} where it was proved to
yield the correct numerical value of the integrals, provided that a weak
assumption concerning the behavior of the gravitational field in the
past ($-\tau \to -\infty$) is satisfied.  This assumption is essentially
that the $\ell$th time derivative of a moment of order $\ell$ tends to a
constant when $-\tau \to -\infty$.  It serves to preclude, for instance,
the emission of a strong burst of radiation in the remote past which
would make a non-physical contribution to the tail integrals (see also
Ref.~\cite{BD88} for a discussion). This assumption is satisfied in the
case where the binary is formed by capture of two bodies moving on an
initial quasi-hyperbolic orbit with small enough energy.  The method
consists (i) in substituting into the tail integrals in Eq.~(\ref{eq:4.3a})
the components of the moments calculated for a {\it fixed} (non decaying)
{\it circular} orbit whose constant orbital frequency is equal to the
{\it current} value of the frequency at time $T_R$, i.e. $\omega \equiv
\omega_{\rm 2PN} = \omega_{\rm 2PN} (T_R)$.  Then it consists (ii) in
evaluating each resulting integral by means of the formula
\begin{equation}
\int^{+\infty}_0 d\tau\,\ln \left({\tau\over 2b}\right) e^{i\Omega\tau} =
 {-1\over \Omega} \left[ {\pi\over 2} {\rm sign} (\Omega)
  + i(\ln (2|\Omega|b) +C)\right] \ , \label{eq:4.6}
\end{equation}
where $C=0.577\cdots$ is Euler's constant and $\Omega$ is the frequency
of the radiation (i.e. a real number $\Omega=\pm n\omega$; ${\rm sign}
(\Omega)\equiv\pm 1$ and $|\Omega|\equiv n\omega )$. Note that this
formula is to be applied even though the left-hand-side of
(\ref{eq:4.6}) is an absolutely convergent integral only when $\Omega$
possesses a strictly positive imaginary part, $Im (\Omega) >0$ (compare
the appendix~A and Sect.~3 in Ref.~\cite{BS93}).  As proved in the
appendix~B of Ref.~\cite{BS93}, the method (i)-(ii) is valid for an
orbit which is actually decaying, and thus for which $\omega(T_R-\tau)$
tends formally to zero when $-\tau\to -\infty$.  The numerical errors
made in following (i)-(ii) have been shown to be of order $O[\xi\ln
\xi]$, where $\xi$ is the usual adiabatic small parameter describing the
decay of the orbit $(\xi \sim \dot\omega/\omega^2)$, taken at the
{\it current} instant $T_R$, i.e.  $\xi =\xi (T_R)$. The proof
consists in showing that an adequately defined ``remote
past'' contribution to the tail integrals is itself of order
$O[\xi(T_R)]$, so that only the ``recent'' values of the frequency, near
the current value $\omega =\omega (T_R)$, contribute to the tail
integrals.  Although this has not rigorously been shown, the proof could
in principle be extended to orbits which had a non negligible
eccentricity in the past, and had in fact whatever behavior in the past
which is consistent with the weak assumption made above.  Note that the
decay of the orbit is driven by radiation reaction effects which are of
order $O(\varepsilon^5)$ in the post-Newtonian parameter $\varepsilon$,
and so the adiabatic parameter $\xi$ is itself of order
$O(\varepsilon^5)$.  One can therefore safely neglect in the 2PN
waveform all the errors brought about by the above procedure (i)-(ii).

We now calculate the two independent polarizations associated with the tail
contribution (\ref{eq:4.3}) with respect to two unit directions ${\bf
P}$ and ${\bf Q}$ perpendicular to ${\bf N}$ and such that ${\bf N},$
${\bf P}$, ${\bf Q}$ forms a right-handed orthonormal triad. We adopt the
usual convention that ${\bf P}$ and ${\bf Q}$ lie along the major and
minor axes of the projection of the circular orbit on the plane of the
sky, respectively, and we denote by $c=\cos i$ and $s=\sin i$ the cosine
and sine of the angle between the line of sight ${\bf N}$ and the normal
to the orbital plane ($c=\cos i$ is not to be confused with the speed of
light which we denote exceptionally by $c_0$ in
Eqs.~(\ref{eq:4.7})-(\ref{eq:4.10})).
Denoting by $\phi$ the instantaneous orbital phase of the binary
(defined as an angle, oriented in the sense of the motion, such that
$\phi ={\pi\over 2}$ (mod $2\pi$) when the relative direction of the two
bodies is ${\bf n}={\bf P}$), and using the relevant time-derivatives
and contractions of {\it Newtonian} moments as readily calculated from
Appendix~\ref{sec:apc}, we follow the items (i)-(ii) above and bring the
two ``plus'' and ``cross'' polarizations $(h_+)_{\rm tail} \equiv
{1\over 2} (P_iP_j -Q_iQ_j)$ $(h^{TT}_{ij})_{\rm tail}$ and
$(h_\times)_{\rm tail} \equiv{1\over 2}(P_iQ_j +Q_iP_j) (h^{TT}_{ij})_
{\rm tail}$ into the form
\begin{eqnarray}
 (h_+)_{\rm tail}&=&{4G^4\over c_0^7R} {\nu m^4\over r^4} \int^{+\infty}_0
 d\tau\left\{ -4(1+c^2)\ln \left( {\tau\over 2b_1}\right) \cos
 [2\phi (T_R-\tau)]\right. \nonumber \\
 &&\qquad +\gamma^{1/2} (X_2-X_1)s\left[{81\over 8} (1+c^2) \ln
 \left({\tau\over 2b_2}\right)\sin[3\phi(T_R-\tau)]\right.\nonumber\\
 &&\qquad +{1\over 8} (5+c^2) \ln \left( {\tau\over 2b_2}\right)
 \sin [\phi (T_R-\tau)]
 \left.\left. - {3\over 10} \sin [\phi (T_R-\tau)]\right]
 \right\}\ , \label{eq:4.7}
\end{eqnarray}
and
\begin{eqnarray}
 (h_\times)_{\rm tail} ={4G^4\over c_0^7R}{\nu m^4\over r^4}\int^{+\infty}_0
 &d\tau&\left\{ -8 c\ln \left( {\tau\over 2b_1}\right) \sin
 [2\phi (T_R-\tau)]\right. \nonumber \\
 &+& \gamma^{1/2} (X_2-X_1)sc\left[-{81\over 4}\ln
  \left( {\tau\over 2b_2}\right) \cos [3\phi (T_R-\tau)]\right.\nonumber\\
 &-& {3\over 4} \ln \left( {\tau\over 2b_2}\right)
  \cos [\phi (T_R-\tau)] \left.\left.
  - {3\over 10} \cos [\phi (T_R-\tau)]\right] \right\}\ . \label{eq:4.8}
\end{eqnarray}
Still following (i)-(ii) we now compute (\ref{eq:4.7})-(\ref{eq:4.8}) by
replacing $\phi$ by a linear function of time, $\phi (T)=\omega T+\phi_0$
where $\omega =\omega (T_R)$ is the current value of the orbital frequency,
and by applying the formula (\ref{eq:4.6}) to each integrals. As a result we
get
\begin{eqnarray}
 (h_+)_{\rm tail} &=& {4G^3\over c_0^7R} {\nu m^3\omega\over r}
 \left\{ 2(1+c^2)\left[ (\ln (4\omega b_1) +C) \sin
  2\phi +{\pi\over 2} \cos 2\phi \right] \right. \nonumber \\
&&+ \gamma^{1/2} (X_2-X_1)s \left[ {27\over 8} (1+c^2) \left[ (\ln
(6\omega b_2)+C)\cos 3\phi -{\pi\over 2}\sin 3\phi \right]\right.\nonumber\\
 &&+{1\over 8} (5+c^2)\left[ (\ln (2\omega b_2)+C)
 \cos \phi -{\pi\over 2} \sin \phi \right]
 \left.\left. + {3\over 10} \cos \phi\right] \right\}\ , \label{eq:4.9}
\end{eqnarray}
and
\begin{eqnarray}
 (h_\times)_{\rm tail} &=& {4G^3\over c_0^7R} {\nu m^3\omega\over r}
 \left\{ -4 c\left[ (\ln (4\omega b_1)+C) \cos
  2\phi -{\pi\over 2} \sin 2\phi \right]\right. \nonumber \\
 && +\gamma^{1/2} (X_2-X_1)sc\left[ {27\over 4}\left[ (\ln
 (6\omega b_2)+C)\sin 3\phi +{\pi\over 2}\cos 3\phi\right]\right.\nonumber\\
 && +{3\over 4} \left[ (\ln (2\omega b_2) +C)
 \sin \phi + {\pi\over 2} \cos \phi \right]
 \left.\left.+ {3\over 10} \sin \phi \right]
 \right\}\ . \label{eq:4.10}
\end{eqnarray}
The orbital frequency $\omega$ and phase $\phi$ in
(\ref{eq:4.9})-(\ref{eq:4.10})
denote the current values $\omega =\omega (T_R)$ and $\phi =\phi (T_R)$.
Both $\omega$ and $\phi$ will be computed in the next subsection (see
Eqs.~(\ref{eq:4.28})-(\ref{eq:4.30})). [Evidently, though during the
computation of Eqs.~(\ref{eq:4.9})-(\ref{eq:4.10}), $\phi$ has been
replaced, following Ref.~\cite{BS93}, by a linear function of time, its
actual time variation is nonlinear.]

\subsection{The energy loss and the associated binary's orbital phasing}
\label{sec:4.2}

 The 2PN-accurate energy loss given by Eq.~(\ref{eq:2.5}) is split,
similarly to the waveform, into an ``instantaneous" contribution and a
``tail'' one. Let us deal first with the instantaneous contribution,
which is defined by
\begin{eqnarray}
\left( {dE_B\over dT_R}\right)_{\rm inst} = - {G\over c^5} \biggl\{
  {1\over 5} I^{(3)}_{ij} I^{(3)}_{ij}
 &+&{1\over c^2} \left[ {1\over 189} I^{(4)}_{ijk} I^{(4)}_{ijk}
 + {16\over 45} J^{(3)}_{ij} J^{(3)}_{ij}\right] \nonumber \\
 &+&{1\over c^4} \left[ {1\over 9072} I^{(5)}_{ijkm} I^{(5)}_{ijkm}
 + {1\over 84} J^{(4)}_{ijk} J^{(4)}_{ijk}\right] \biggr\}\ .
 \label{eq:4.11}
\end{eqnarray}
All the needed moments to compute this contribution have been given in
Eqs.~(\ref{eq:4.4}). Their relevant time-derivatives and ``squares"
can be found in Appendix~C. By a quite straightforward computation
we get
\begin{equation}
\left({dE_B\over dT_R}\right)_{\rm inst}=-\frac{32}{5}\frac{c^5}{G}
\nu^2\gamma^5
\left\{1\;-\;\gamma \left( {2927\over 336} + {5\over 4}\nu \right) +
 \gamma^2 \left( {293383\over 9072} + {380\over 9}\nu \right) \right\}
     \ . \label{eq:4.12}
\end{equation}
 As for the tail contribution, it reads
(when retaining only the terms which contribute to the 2PN order)
\begin{eqnarray}
 \left({dE_B\over dT_R}\right)_{\rm tail}&=&-{2G\over 5c^5}
 {2Gm\over c^3} I^{(3)}_{ij} (T_R) \int^{+\infty}_0 d\tau\, \ln
 \left( {\tau\over 2b_1}\right) I^{(5)}_{ij} (T_R-\tau)\ . \label{eq:4.13}
\end{eqnarray}
Note that at the 2PN order there appears only the tail integral associated
with the mass {\it quadrupole} moment $I_{ij}$, which can be replaced by its
usual {\it Newtonian} expression. We now evaluate Eq.~(\ref{eq:4.13}) by
the same method (i)-(ii) as used previously for the tails in the
waveform.  Namely we replace in (\ref{eq:4.13}) the third and fifth
time-derivatives of the quadrupole moment by Newtonian quantities valid
for circular orbits, and we perform explicitly the contractions between
these moments, being careful that one of them is taken at the current
instant $T_R$ while the other is taken at the former instant $T_R-\tau$.
This readily brings (\ref{eq:4.13}) into the form
\begin{equation}
 \left( {dE_B\over dT_R} \right)_{\rm tail} = {512G^6\over 5c^8}
 {\nu^2m^7\over r^8} \int^{+\infty}_0 d\tau\, \ln \left( {\tau\over
  2b_1}\right)
  \cos [2\phi (T_R-\tau)]\ , \label{eq:4.14}
\end{equation}
where for this computation $\phi$ can be assumed to be a linear phase
\cite{BS93}.
Then the use of the integration formula (\ref{eq:4.6}) immediately yields
\begin{equation}
 \left( {dE_B\over dT_R} \right)_{\rm tail} = - {32c^5\over 5G}
 \nu^2\gamma^5 \{ 4\pi \gamma^{3/2}\}\ . \label{eq:4.15}
\end{equation}
There is no dependence on the constant $b$ parametrizing the freedom
in constructing the radiative coordinate system. This is not surprising
because the constant $b$ (or rather $b_1$) enters a term which is a
total derivative (see Eq.~(\ref{eq:4.13})) and thus, as we already noticed
\cite{N1}, which vanishes identically for circular orbits.

 A central result of this paper, namely the complete 2PN-accurate
gravitational energy loss rate from a compact binary moving on a circular
orbit, is now obtained by adding Eqs.~(\ref{eq:4.12}) and (\ref{eq:4.15}):
\begin{eqnarray}
  {dE_B\over dT_R}&=&-{32c^5\over 5G} \nu^2\gamma^5 \left\{ 1 - \gamma
\left( {2927\over 336}+ {5\over 4}\nu\right) + 4\pi\gamma^{3/2}
 + \gamma^2\left( {293383\over 9072} +{380\over 9}\nu
\right) \right\} \ .  \label{eq:4.16}
\end{eqnarray}
The $\gamma$-parameter is $\gamma\equiv Gm/(rc^2)$ where $r$ is the
harmonic radial coordinate (see Eqs.~(\ref{eq:3.10})-(\ref{eq:3.13})).
It is to be noticed that in the form (\ref{eq:4.16}), i.e.  when the
post-Newtonian expansion is parametrized using $\gamma$, there is no
term proportional to $\nu^2$ in the relative 2PN contribution. This fact is
somewhat surprising because all separate pieces making up the energy
loss, i.e. all the different ``squares" of moments listed in Appendix~C,
do contain terms proportional to $\nu^2$. However the final coefficient of
$\nu^2$ in the 2PN correction term turns out to be zero. The expression of
the energy loss (\ref{eq:4.16}) completes several previous investigations
having obtained either the lower-order PN corrections or the limiting
case where the mass of one body is negligible as compared to the other
one (limiting case $\nu\to 0)$. The 0PN leading term in Eq.~(\ref{eq:4.16})
was known from Ref.~\cite{LL0}, and the 1PN correction was added in
Refs.~\cite{WagW76,Gal80}.  The 1.5PN tail correction (4$\pi$ term) was
first computed in the limit $\nu\to 0$ \cite{P93} and then shown to be
also valid for arbitrary mass ratios in Refs.~\cite{BS93,Wi93} based on
Ref.~\cite{BD92}.  [The generalization of the energy loss expression for
non-circular (eccentric) orbits has been obtained in Ref.~\cite{PeM63}
for the 0PN leading term and in Ref.~\cite{BS89} for the 1PN corrections.
Note that for non-circular orbits the tail term $4\pi \gamma^{3/2}$ has
simply to be multiplied by a function $\varphi (e)$ of the eccentricity
of the orbit, but unfortunately this function probably does not admit a
closed analytical form (see Ref.~\cite{BS93} for the numerical graph of
this function).] Finally the 2PN correction term in Eq.~(\ref{eq:4.16})
was known, up to now, only in the limiting case $\nu\to 0$, where it was
computed first numerically \cite{TNaka94} and then analytically
\cite{TSasa94}.  For comparison with the latter references, and for
later convenience, let us re-express the energy loss (\ref{eq:4.16}) in
terms of a new post-Newtonian parameter defined by $x\equiv
(Gm\omega_{\rm 2PN}/c^3)^{2/3}$.  The $\gamma$-parameter is related to
the $x$-parameter by the inverse of Eq.~(\ref{eq:3.11}), namely
\begin{equation}
  \gamma = x \left[ 1+ \left( 1-{1\over 3}\nu\right) x + \left( 1 -
  {65\over 12}\nu\right) x^2\right]\   \label{eq:4.17}
\end{equation}
(which also does not involve $\nu^2$-terms). Inserting
(\ref{eq:4.17}) into (\ref{eq:4.16}) (and keeping consistently
all terms up to the 2PN order) readily yields
\begin{equation}
 {dE_B\over dT_R} = - {32c^5\over 5G} \nu^2x^5 \left\{ 1 + \left( -
 {1247\over 336} - {35\over 12}\nu \right) x + 4\pi\,
   x^{3/2} + \left( -{44711\over 9072} + {9271\over 504}
  \nu + {65\over 18} \nu^2 \right) x^2 \right\} \ . \label{eq:4.18}
\end{equation}
This expression (which we notice involves now a $\nu^2$-term) relates
the two coordinate-independent quantities $dE_B/dT_R$ and $\omega =
\omega_{2PN}$, and is therefore the same in all coordinate systems.
It may thus be compared directly with the expression obtained in
Refs.~\cite{TNaka94,TSasa94} in the limit $\nu\to 0$. We find that the
coefficient $-44711/9072$ agrees with the latter references \cite{N3}.

Let us now denote the 2PN-accurate Bondi energy loss rate
(\ref{eq:4.18}), or total gravitational luminosity, by ${\cal L}_B
\equiv -dE_B/dT_R$, and let us equate, by a standard argument, the rate
of decrease of the {\it dynamical} energy $\cal E$ of the binary system
to the opposite of ${\cal L}_B$, i.e.
\begin{equation}
  {d {\cal E} \over dT_R} = - {\cal L}_B\ . \label{eq:4.19}
\end{equation}
We shall admit here the validity of the balance equation (\ref{eq:4.19})
to the 2PN order, although it has been validated only to the leading 0PN
order and in the case of the binary pulsar\cite{DD81a}. [Note that the 2PN
order is the last order at which the binary admits a conserved energy
$\cal E$ (and a conserved angular momentum ${\cal J}$).] The balance
equation (\ref{eq:4.19}) drives the variations with time of the
instantaneous orbital frequency $\omega$ and phase $\phi$ of the
(quasi-circular) decaying orbit.  These $\omega$ and $\phi$ are the ones
which enter the expression of the waveform (see e.g.
Eqs.~(\ref{eq:4.9})-(\ref{eq:4.10})).  To compute $\omega$ and $\phi$
one must evidently specify first what is the left-hand-side of
Eq.~(\ref{eq:4.19}), i.e.  one must know the 2PN-accurate expression of
the center-of-mass energy ${\cal E}$ for a {\it fixed} (non-decaying)
circular orbit as a function of the parameter $x$.

Let us recall from Ref.~\cite{DS88} (extending Ref.~\cite{DD85})
that the 2PN motion of a binary system moving on an eccentric orbit
admits the following representation:
\begin{mathletters}
 \label{eq:4.20}
\begin{eqnarray}
 n(t-t_0) &=& u-e_t \sin u + {f\over c^4} \sin v + {g\over c^4} (v-u)\ ,
  \label{eq:4.20a}\\
 r &=& a_r (1-e_r \cos u)\ , \label{eq:4.20b}\\
 \phi - \phi_0 &=& K \left\{ v + {F\over c^4} \sin 2v + {G\over c^4}
  \sin 3v \right\}\ , \label{eq:4.20c}
\end{eqnarray}
where
\begin{equation}
 v = 2 \arctan \left[ \left( {1+e_\phi \over 1-e_\phi} \right)^{1/2}
   \tan\, {u\over 2} \right] \ . \label{eq:4.20d}
\end{equation}
\end{mathletters}
The time $t$, separation $r$ between the two bodies and polar angle (or
phase) $\phi$ in these equations correspond to harmonic coordinates.
The motion is parametrized by some ``eccentric'' and ``real'' anomalies
$u$ and $v$, and ten constants enter Eqs.~(\ref{eq:4.20}) besides the
initial $t_0$ and $\phi_0$:  $n\equiv 2\pi/P$ where $P$ is the time of
return to the periastron (or period);  $K\equiv \Theta /2\pi$ where
$\Theta$ is the angle of return to the periastron $(K-1$ is the relative
periastron advance per rotation);  the semi-major axis $a_r$;  three
types of eccentricities $e_t$, $e_r$ and $e_\phi$;  and four constants
$f$, $g$, $F$ and $G$ entering purely 2PN terms.  Most important for our
purpose are the expressions of the constants $n$ and $K$ obtained in
Ref.~\cite{DS88} (see Eqs.~(\ref{eq:3.11})-(\ref{eq:3.12}) there) in
terms of the constant center-of-mass energy ${\cal E}$ and angular
momentum ${\cal J}$.  Denoting ${\cal E} =\nu m\overline{\cal E}$ and
${\cal J} =G\nu m^2h$ we have
\begin{eqnarray}
 n&=&{(-2\overline{\cal E})^{3/2}\over Gm} \left\{ 1 +{1\over 4} (15-\nu)
{\overline{\cal E}\over c^2} +
 {3\over 32} \left( 185+10\nu +{11\over 3} \nu^2\right)
{\overline{\cal E}^2\over c^4}
- {3\over 2} (5-2\nu) {(-2\overline{\cal E})^{3/2}\over c^4h}\right\}\ ,
 \nonumber \\
\label{eq:4.21} \\
 K&=&1+{3\over c^2h^2} \left\{ 1+ \left( {5\over 2}-\nu\right)
 {\overline{\cal E}\over c^2}
 + \left( {35\over 4} -{5\over 2}\nu \right) {1\over c^2h^2} \right\} \ .
 \label{eq:4.22}
\end{eqnarray}
The other constants in Eqs.~(\ref{eq:4.20}) have been computed in
Ref.~\cite{SW93}; however we shall here only need, in addition to
Eqs.~(\ref{eq:4.21}) and (\ref{eq:4.22}), the fact that in the
{\it circular} orbit case, which is defined by $e_r=0$, the two other
eccentricities $e_t$ and $e_\phi$ vanish to 2PN order, hence the
two anomalies $u$ and $v$ agree to this order, and that the three terms
involving the constants $f$, $F$ and $G$ also vanish (see
\cite{SW93}). The 2PN-accurate circular motion of the binary is
thus simply described by the equations
\begin{mathletters}
\label{eq:4.23}
\begin{eqnarray}
 r &=& a_r \ , \label{eq:4.23a}\\
 \phi - \phi_0 &=& nK (t- t_0)\ , \label{eq:4.23b}
\end{eqnarray}
\end{mathletters}
which show that the orbital frequency $\omega$ in the circular orbit case
is equal to the product $nK$. Now $\omega=nK$ can straightforwardly be
obtained from Eqs.~(\ref{eq:4.21})-(\ref{eq:4.22}) and the fact that
$\overline{\cal E}$ and $h$ are related to each other by the 1PN-accurate
relation $1+2 \overline{\cal E}h^2 = {1\over 2} (9+\nu) \overline{\cal E}
/c^2$ (consequence of $e_r =0$). We find
\cite{N4}
\begin{equation}
 \omega = {(-2\overline{\cal E})^{3/2}\over Gm} \left\{ 1 - {1\over 4}
   (9+\nu) {\overline{\cal E}\over c^2}
  + {3\over 32} \left( 297 - 134\nu + {11\over 3} \nu^2\right)
  {\overline{\cal E}^2\over c^4} \right\} \ . \label{eq:4.24}
\end{equation}
The inversion of Eq.~(\ref{eq:4.24}) then yields our desired relation
between the energy ${\cal E}=\nu m\overline{\cal E}$ and the parameter
$x=(Gm\omega /c^3)^{2/3}$ for circular orbits:
\begin{equation}
 {\cal E} = -{c^2\over 2} \nu m x \left\{ 1 - {1\over 12} (9+\nu) x -
  {1\over 8} \left( 27 - 19\nu +{\nu^2\over 3}\right) x^2 \right\}\ .
 \label{eq:4.25}
\end{equation}

 With Eqs.~(\ref{eq:4.18}) and (\ref{eq:4.25}) in hand, it is now a
simple matter to transform the energy balance equation (\ref{eq:4.19})
into the ordinary differential equation
\begin{equation}
  d\theta = {1\over 64} dx\, x^{-5} \left\{ 1 + \left( {743\over 336}
   + {11\over 4} \nu \right) x - 4\pi x^{3/2}
  + \left({3058673\over 1016064}+{5429\over 1008}\nu
  + {617\over 144} \nu^2 \right) x^2 \right\} \ , \label{eq:4.26}
\end{equation}
where we have introduced for convenience the adimensional time variable
\begin{equation}
 \theta \equiv {c^3\nu\over 5Gm} \, T_R\ . \label{eq:4.27}
\end{equation}
Solving (\ref{eq:4.26}) leads to the variation in time $\theta$ of the
parameter $x$ and hence of the instantaneous frequency $\omega$ of the
quasi-circular orbit. Similarly, integrating $d\phi =\omega dT_R = (5/\nu)
x^{3/2} d\theta$ leads to the variation of the instantaneous orbital
phase $\phi$.  We finally obtain
\begin{eqnarray}
 {Gm\omega\over c^3} &=& {1\over 8} (\theta_c -\theta)^{-3/8}
 \left\{ 1 + {1\over 8}p
  (\theta_c-\theta)^{-1/4}-{3\pi\over 10}
  (\theta_c -\theta)^{-3/8}
  + {1\over 64} q  (\theta_c-\theta)^{-1/2}\right\}\ ,
         \label{eq:4.28}\\
 \phi_c-\phi &=& {1\over \nu} (\theta_c -\theta)^{5/8}
 \left\{ 1 + {5\over 24} p
  (\theta_c-\theta)^{-1/4}-{3\pi\over 4}
  (\theta_c -\theta)^{-3/8}
                 +{5\over 64}q
(\theta_c-\theta)^{-1/2}\right\}\ ,\label{eq:4.29}
\end{eqnarray}
where
\begin{equation}
 p \equiv {743\over 336} + {11\over 4} \nu \ , \qquad
 q \equiv {1855099\over 225792} + {56975\over 4032} \nu
  + {371\over 32} \nu^2\ , \label{eq:4.30}
\end{equation}
and where $\phi_c$ and $\theta_c$ denote the values of the phase and
adimensional time (\ref{eq:4.27}) at the instant of the coalescence. The
frequency and phase (\ref{eq:4.28})-(\ref{eq:4.29}) have to be inserted
into the expression of the waveform. Note that the coefficient of the 2PN
contribution in Eq.~(\ref{eq:4.29}) (proportional to $q$), increases by
52\% between the test-mass limit ($\nu =0)$ and the equal-mass case
$(\nu =1/4)$.  This shows that finite mass effects (which cannot be
obtained in perturbation calculations of black hole spacetimes) play a
very significant role in the definition of 2PN-accurate theoretical
waveforms.  This proves the importance of post-Newtonian generation
formalisms for constructing accurate templates to be used in matched
filtering of the data from future gravitational-wave detectors.

\acknowledgments
One of us (BRI) would like to acknowledge the hospitality of IHES during
the initial phase of this collaboration.

\appendix
\section{The mass monopole and dipole moments}
\label{sec:apa}

Since the expression (\ref{eq:2.17}) of the 2PN-accurate mass multipole
moment is valid for all orders of multipolarity $\ell$, it is
important to verify that for the lowest orders $\ell =0$ and $\ell =1$
it reduces to the expressions of the {\it conserved} mass monopole and
mass dipole moments. This verification has already been done in
Appendix~B of \cite{B94} in the mass monopole case $\ell=0$. Here we
shall check that for two bodies moving on a circular orbit the mass
monopole reduces to the expression computed in \cite{DD81b}, and the mass
dipole is {\it zero}, as it must be since we are using a mass-centred
frame.

The mass monopole $I$ and dipole $I_i$, which are obtained by setting
$\ell =0$ and $\ell=1$ in Eq.~(\ref{eq:2.17}), are split into three
``Compact'', ``Y'', and ``W'' contributions according to
Eq.~(\ref{eq:3.1}) and evaluated separately using the method developed
in Sect.~\ref{sec:3}. The ``W'' contributions have in fact already been
computed in Eqs.~(\ref{eq:3.70}) and (\ref{eq:3.71a})-(\ref{eq:3.71b}).
We quote only the results. For $\ell =0$ we find
\begin{mathletters}
\label{eq:A1}
\begin{eqnarray}
  I^{(C)} &=& m \left[ 1 -{1\over 2}\nu\, \gamma - {1\over 8}\nu (1-15\nu)
     \gamma^2 \right]\ , \label{eq:A1a}\\
  I^{(Y)} &=& 2m\nu (1-\nu)\, \gamma^2\ , \label{eq:A1b}\\
  I^{(W)} &=& -m\nu\, \gamma^2\ . \label{eq:A1c}
\end{eqnarray}
\end{mathletters} Adding together these contributions leads to
\begin{equation}
 I = m \left[ 1 - {1\over 2} \nu\,\gamma + {1\over 8}\nu (7-\nu)
      \gamma^2 \right]\ ,  \label{eq:A2}
\end{equation}
which is in agreement with Eq.~(20) in Ref.~\cite{DD81b} when specialized
to circular orbits. For $\ell=1$ we get
\begin{mathletters}
\label{eq:A3}
\begin{eqnarray}
  I^{(C)}_i &=& (X_2-X_1) m\nu \,\gamma^2 \left( {2\over 7}
         + {29\over 35}\nu\right) x^i\ , \label{eq:A3a}\\
  I^{(Y)}_i &=& (X_2-X_1) m\nu \,\gamma^2 \left(-{9\over 7}
         - {29\over 35}\nu\right) x^i\ , \label{eq:A3b}\\
  I^{(W)}_i &=& (X_2-X_1) m\nu\,\gamma^2\, x^i\ , \label{eq:A3c}
\end{eqnarray}
\end{mathletters}
in which we have used (for the ``C'' term only) the 2PN-accurate mass-centred
frame equation (\ref{eq:3.7}). The three contributions (\ref{eq:A3})
sum up to zero,
\begin{equation}
  I_i = 0\ , \label{eq:A4}
\end{equation}
as was to be checked.

\section{Alternative derivation of the $W$-term}
\label{sec:apb}

We have computed in Sect.~\ref{sec:3.3} of the main text the cubically
nonlinear ``$W$'' term (defined by Eq.~(\ref{eq:3.30})) for all values of
the order of multipolarity $\ell$, but in the special case where the source
is a {\it binary} system. In this appendix we present an alternative
derivation of this $W$-term which is valid for a general fluid system
(and hence for a system made of $N$ compact bodies), but is limited to
the {\it quadrupole} case $\ell =2$. When $N=2$ {\it and} $\ell=2$, i.e.
when one evaluates the $W$-term for a binary system {\it and} in the
quadrupole case ---~this
is what interests us in this paper~---, we find that both
derivations agree on the result given by Eq.~(\ref{eq:3.72}).

Specializing the definition (\ref{eq:3.30}) to $\ell=2$, we thus want to
compute
\begin{equation}
  I^{(W)}_{ij} = {1\over \pi Gc^4} {\rm FP}_{B=0} \int d^3 {\bf x}
|{\bf x}|^B \hat x_{ij} W_{km} \partial_{km} U\ , \label{eq:B1}
\end{equation}
where we recall that the potential $W_{km}$ is defined by
\begin{equation}
 W_{km} ({\bf x},t) = - {1\over 4\pi} \int {d^3{\bf x}'\over |{\bf
x}-{\bf x}'|} [\partial_k U \partial_m U] ({\bf x}',t)\ . \label{eq:B2}
\end{equation}
We first perform on (\ref{eq:B1}) two integrations by parts in order to
shift the spatial derivatives acting on $U$ to the left side of the
integrand.  As can be proved thanks to Eq.~(\ref{eq:4.2}) of
Ref.~\cite{B94}, all the terms coming from the differentiation of the
analytic continuation factor $|{\bf x}|^B$ and having explicitly $B$ as
a factor are zero. [Indeed, recall that Eq.~(\ref{eq:4.2}) of
Ref.~\cite{B94} permits one to freely integrate by parts all terms in the
source moment (\ref{eq:2.17}) as if the analytic continuation factor and
the ``finite part'' were absent.] The surface terms are also zero by
analytic continuation.  Hence we can split the $W$-term into two pieces,
\begin{equation}
 I^{(W)}_{ij} = I^{(W1)}_{ij} + I^{(W2)}_{ij}\ ,\label{eq:B3}
\end{equation}
given respectively by
\begin{eqnarray}
 I^{(W1)}_{ij} &=& {2\over \pi Gc^4}\, {\rm STF}_{ij}\ {\rm FP}_{B=0}
 \int d^3 {\bf x} |{\bf x}|^B W_{ij} U\ , \label{eq:B4}\\
 I^{(W2)}_{ij} &=& {1\over \pi Gc^4} {\rm FP}_{B=0} \int d^3 {\bf x}
  |{\bf x}|^B [4\partial_k W_{k<i} x_{j>} + \partial_{km} W_{km}
   \hat x_{ij} ] U\  \label{eq:B5}
\end{eqnarray}
(where the angular brackets $<\,>$ denote the tracefree projection). In
(\ref{eq:B5}) the divergences of the potential $W_{ij}$ are given by
\begin{mathletters}
\label{eq:B6}
\begin{eqnarray}
 \partial_k W_{ik} &=& {1\over 4} \partial_i (U^2) + {G\over 2} \int
 {d^3{\bf x'}\over |{\bf x}-{\bf x}'|} [\sigma \partial_i U - U\partial_i
  \sigma] ({\bf x}',t)\ , \label{eq:B6a} \\
 \partial_{km} W_{km} &=& {1\over 4} \Delta (U^2) + {G\over 2} \int
 {d^3{\bf x'}\over |{\bf x}-{\bf x}'|} [\sigma \Delta U - U\Delta
  \sigma] ({\bf x}',t)\ . \label{eq:B6b}
\end{eqnarray}
\end{mathletters}

 The most interesting contribution to compute is the first one, i.e.
$I^{(W1)}_{ij}$ given by (\ref{eq:B4}). This contribution
has no explicit dependence on ${\bf x}$ in the integrand; this fact, which is
special to the quadrupolar case, allows the alternative derivation
followed in this appendix. To compute $I_{ij}^{(W1)}$ we employ
the kernel $g({\bf x}; {\bf y}_1, {\bf y}_2)$ already introduced
in Eqs.~(\ref{eq:3.42}),
\begin{equation}
g({\bf x};{\bf y}_1, {\bf y}_2) = \ln (r_1 + r_2 + r_{12})\ , \label{eq:B7}
\end{equation}
which satisfies
\begin{equation}
 \Delta_x g = {1\over r_1 r_2}  \label{eq:B8}
\end{equation}
in the sense of distribution theory. (We denote $r_1=|{\bf x}-{\bf y}_1|$,
$r_2=|{\bf x}-{\bf y}_2|,$ $r_{12}=|{\bf y}_1-{\bf y}_2|.$) The
essence of the present computation of $I^{(W1)}_{ij}$ is to consider
a secondary kernel $f$ defined by
\begin{eqnarray}
 f({\bf x};{\bf y}_1,{\bf y}_2) &=& {1\over 3} ({\bf r}_1\cdotp {\bf r}_2)
  \left[ \ln (r_1 + r_2 + r_{12}) - {1\over 3} \right] + {1\over 6}
    (r_{12} r_1 + r_{12} r_2 - r_1 r_2) \ ,\label{eq:B9}
\end{eqnarray}
and whose main property is to satisfy
\begin{equation}
 \Delta_x f = 2\,g \label{eq:B10}
\end{equation}
in the sense of distributions. This property is easily checked on the
expression (\ref{eq:B9}).  We now recall that the potential $W_{ij}$ can
be expressed in terms of the kernel $g$ as
\begin{equation}
  W_{ij} ({\bf x},t) = G^2 \int\!\!\int d^3 {\bf y}_1d^3{\bf y}_2 \sigma
 ({\bf y}_1,t) \sigma ({\bf y}_2,t) {\partial^2\over
 \partial y^{i}_1 \partial y^{j}_2} \{ g({\bf x};{\bf y}_1, {\bf y}_2)\}
      \ . \label{eq:B11}
\end{equation}
Thus we see that a secondary potential defined by the same expression
as (\ref{eq:B11}) but with $g$ replaced by the secondary kernel $f$,
i.e.
\begin{equation}
  w_{ij} ({\bf x},t) = G^2 \int\!\!\int d^3 {\bf y}_1d^3{\bf y}_2 \sigma
 ({\bf y}_1,t) \sigma ({\bf y}_2,t) {\partial^2\over
 \partial y^{i}_1 \partial y^{j}_2} \{ f({\bf x};{\bf y}_1, {\bf y}_2)\}
      \ , \label{eq:B12}
\end{equation}
necessarily satisfies
\begin{equation}
 \Delta  w_{ij} = 2\,  W_{ij} \label{eq:B13}
\end{equation}
in the sense of distributions.
Substituting (\ref{eq:B13}) into (\ref{eq:B4}) leads to
\begin{equation}
 I^{(W1)}_{ij} = {1\over \pi Gc^4}\, {\rm STF}_{ij}\, {\rm FP}_{B=0} \int
 d^3 {\bf x} |{\bf x}|^B U \Delta w_{ij}\ .\label{eq:B14}
\end{equation}
The next step is to integrate by parts the Laplace operator in the
integrand of (\ref{eq:B14}). However, note that it is {\it a priori} not
possible to perform this integration by parts ignoring the analytic
continuation factor $|{\bf x}|^B$ and the finite part symbol. For instance,
the result (4.2) of Ref.~\cite{B94} invoked above does not apply to this
case.  We thus integrate by parts the integral (B14) keeping
all terms coming from the differentiation of $|{\bf x}|^B$.  This yields
(using $\Delta U =-4\pi G\sigma$)
\begin{equation}
 I^{(W1)}_{ij} = - {4\over c^4} {\rm STF}_{ij} \int d^3 {\bf x} \sigma
 w_{ij} + R_{ij}\ , \label{eq:B15}
\end{equation}
where the first term has a compact support (we have removed on it the
analytic continuation factor and the finite part symbol), and
where the second term is explicitly given by
\begin{equation}
 R_{ij} = {1\over \pi Gc^4}\, {\rm STF}_{ij}\, {\rm FP}_{B=0} \int d^3
  {\bf x} |{\bf x}|^{B-2} w_{ij} \Bigl[B(B+1) + 2Bx^k\partial_k\Bigr] U\ .
  \label{eq:B16}
\end{equation}
An equivalent expression for $R_{ij}$ which is convenient for our purpose
is easily obtained by substitution
of the expression (\ref{eq:B12}) of the potential $w_{ij}$ and use of
$U=\int d^3{\bf y}_3 \sigma ({\bf y}_3,t) / |{\bf x}-{\bf y}_3|$. We have
\begin{equation}
 R_{ij} = {G^2\over \pi c^4}\,{\rm STF}_{ij} \int\!\!\!\int\!\!\!\int
  d^3{\bf y}_1 d^3{\bf y}_2 d^3{\bf y}_3 \, \sigma (y_1) \sigma (y_2)
 \sigma (y_3)\, {\partial^2\over \partial y^i_1
  \partial y^j_2} \{ K ({\bf y}_1, {\bf y}_2; {\bf y}_3)\} \label{eq:B17}
\end{equation}
where
\begin{equation}
 K({\bf y}_1, {\bf y}_2 ; {\bf y}_3) = {\rm FP}_{B=0}\left[ B(B-1)-2B y^k_3
 {\partial\over \partial y^k_3} \right] \left\{ \int d^3 {\bf x}
 |{\bf x}|^{B-2} {f({\bf x};{\bf y}_1, {\bf y}_2)\over |{\bf x}-{\bf y}_3|}
 \right\}\ . \label{eq:B18}
\end{equation}
To compute $K({\bf y}_1, {\bf y}_2 ; {\bf y}_3)$, we need to control the
{\it pole part} when $B\to 0$ of the integral on the right-hand side of
(\ref{eq:B18}). This is because of the explicit factors $B$ and $B^2$ in
front of the integral.  [Note that it is important to keep the factors
$B$ {\it and} $B^2$ in front since as we shall see the integral involves
both a simple {\it and} a double pole when $B\to 0$.] The pole part of
the integral in (\ref{eq:B18}) depends only on the behavior of the
integrand at the upper bound $|{\bf x}| \to \infty$, so we need only to
consider the asymptotic expansion of the kernel $f({\bf x};  {\bf y}_1,
{\bf y}_2)$ when $|{\bf x}| \to \infty$, or equivalently when ${\bf
y}_1, {\bf y}_2 \to 0$.  It is then easily shown that the only terms in
the latter expansion of $f$ which generate poles when $B\to 0$ are either
of the type a regular solution of Laplace's equation, i.e.  $\hat x_L$
times a function of ${\bf y}_1$ and ${\bf y}_2$, or of the type $\hat
x_L \ln |{\bf x}|$ times a function of ${\bf y}_1$, ${\bf y}_2$ (note
that the expansion of $f$ involves a logarithm of $|{\bf x}|$).
We shall slightly improperly refer to these terms as the ``harmonic''
terms in $f$.  Their computation can be greatly simplified by noticing
that in the asymptotic expansion of $g=\ln (r_1+r_2+r_{12})$ when ${\bf
y}_1$, ${\bf y}_2 \to 0$, only the three leading order contributions
---~constant, linear, and quadratic in ${\bf y}_1$, ${\bf y}_2$~--- can
contribute to $f_{\rm harmonic}$. Indeed, the higher order contributions,
at least cubic in ${\bf y}_1$, ${\bf y}_2$, necessarily involve a function
of ${\bf x}$ whose dimension is that of $1/|{\bf x}|^n$ with $n\geq 3$
(because $g$ is dimensionless), and thus which will never yield a term of
the type $\hat x_L$ or $\hat x_L \ln |{\bf x}|$ when multiplied by
${\bf r}_1.{\bf r}_2 = {\bf x}^2 - {\bf x}.{\bf y}_1 - {\bf x}.{\bf y}_2
+{\bf y}_1.{\bf y}_2$.  Computing the expansion of $g$ when ${\bf y}_1$,
${\bf y}_2 \to 0$, and then the expansion of $f$, we obtain
\begin{equation}
 f_{\rm harmonic} = {1\over 3}{\bf y}_1.{\bf y}_2\ln (2|{\bf x}|) -{1\over 3}
({\bf x}.{\bf y}_1 + {\bf x}.{\bf y}_2) \left[ \ln (2|{\bf x}|) -{1\over 3}
 \right]\ . \label{eq:B19}
\end{equation}
[This computation of the ``harmonic'' terms in $f$ bears a resemblance to
the computation in Sect.IIIC of the ``homogeneous" part of the function
$W$.] The pole part of the integral in (\ref{eq:B18}) is then
straightforwardly obtained from the replacements $f\to f_{\rm harmonic}$
and $|{\bf x}-{\bf y}_3|^{-1} \to |{\bf x}|^{-1}(1+{\bf y}_3.{\bf x}/
|{\bf x}|^2)$ (we do not need to expand $|{\bf x}-{\bf y}_3|^{-1}$
further since $f_{\rm harmonic}$ involves only terms with multipolarity 0
or 1). We find a simple and a double pole. Then we apply to the latter pole
part the operator $B[B-1 -2 y^k_3 \partial/\partial y^k_3]$ present in
(\ref{eq:B18}) (being careful about correctly handling the double pole),
and compute the finite part at $B=0$.  The result is
\begin{equation}
 K ({\bf y}_1, {\bf y}_2; {\bf y}_3) = {1\over 3} {\bf y}_1 .{\bf y}_2
 \biggl(\ln 2+1\biggr) + {1\over 9} ({\bf y}_1. {\bf y}_3 + {\bf y}_2 .
 {\bf y}_3) \left( \ln 2 -{1\over 3} \right)\ .  \label{eq:B20}
\end{equation}
This is a non zero result; however the quantity of interest in (\ref{eq:B17})
is
\begin{equation}
 {\rm STF}_{ij} {\partial^2\over \partial y^i_1 \partial y^j_2}
 \{ K ({\bf y}_1 , {\bf y}_2 ; {\bf y}_3) \} = 0
  \quad \Longrightarrow \quad R_{ij} = 0\ , \label{eq:B21}
\end{equation}
which shows that the contribution $I_{ij}^{(W1)}$ given by
(\ref{eq:B15}) reduces to its first term, i.e.  to the manifestly
compact support integral
\begin{equation}
 I^{(W1)}_{ij} = - {4\over c^4}\, {\rm STF}_{ij} \int d^3{\bf x}\sigma\,
    w_{ij}\ .  \label{eq:B22}
\end{equation}
This expression is our main result because the second contribution
$I^{(W2)}_{ij}$, given by (\ref{eq:B5}), is evaluated without problem
by substituting into it the formulas (\ref{eq:B6})  giving the
divergences of the potential $W_{ij}$ and using at various places the
function $Y^L$ of Eq.~(\ref{eq:3.23}).  Adding up all the terms
constituting $I^{(W2)}_{ij}$ to the first contribution $I^{(W1)}_{ij}$
given by (\ref{eq:B22}), we arrive finally at the following expression
for the cubically nonlinear quadrupole term:
\begin{eqnarray}
 I^{(W)}_{ij}&=&-{4G^2\over c^4}\,{\rm STF}_{ij}\int\!\!\int\!\!\int
 d^3{\bf y}_1 d^3{\bf y}_2 d^3{\bf y}_3 \sigma (y_1) \sigma (y_2) \sigma
  (y_3)\nonumber\\
 && \times \left\{ {\partial^2\over \partial y^{i}_1 \partial y^{j}_2}
  \left( {1\over 3}{\bf r}_{13} \cdotp {\bf r}_{23} \left[ \ln (r_{13} +
   r_{23} + r_{12}) -{1\over 3}\right]\right.
+ {1\over 6} (r_{12}r_{13} +
   r_{12}r_{23} - r_{13}r_{23})  \right) \nonumber \\
 && \qquad+{1\over 4 r_{13} r_{23}} (y^i_3
  y^{j}_3 - y^{i}_1 y^{j}_1 -y^{i}_2 y^{j}_2)
  - {{\bf r}_{12}\cdotp {\bf r}_{13}\over 6r^3_{13} r_{12}}
 (y^{i}_1 y^{j}_1 + y^{i}_1 y^{j}_2 + y^{i}_2 y^{j}_2) \nonumber\\
 &&\qquad \left. + {r_{12}\over 6r^3_{13}}
  ( 4 y^{i}_1 y^{j}_3 - 4y^{i}_1 y^{j}_1 + 5 y^{i}_2 y^{j}_3
  - 5 y^{i}_1 y^{j}_2 ) \right\}\ . \label{eq:B23}
\end{eqnarray}
This expression is valid for a general fluid system. We can evaluate in
an explicit way the differentiations with respect to ${\bf y}_1$ and
${\bf y}_2$ it contains. In doing so one finds that thanks to the
tracefree projection there is in fact no logarithmic term in
(\ref{eq:B23}). After reduction of (\ref{eq:B23}) to the 2-body case
(using as usual $\delta$-functions and formally discarding all infinite
self-energy terms), we get
\begin{equation}
 I^{(W)}_{ij} = - {G^2m_1m_2\over c^4 r^2_{12}}\, {\rm STF}_{ij}
  \left\{ (5m_1+2m_2) y^{ij}_1 + (5m_2+2m_1) y^{ij}_2
  - 6 (m_1+m_2) y^{i}_1 y^{j}_2 \right\}\ . \label{eq:B24}
\end{equation}
Changing notation with $y^i_{12} \equiv y^i_1 - y^i_2$ finally leads to
\begin{equation}
 I^{(W)}_{ij} = - {G^2m_1m_2\over c^4r^2_{12}}\, {\rm STF}_{ij}
  \left\{ m_1 y^{ij}_1 + m_2 y^{ij}_2 + 2 (m_1+m_2) y^{ij}_{12}
   + 2 (m_1 y^{i}_1- m_2y^{i}_2) y^{j}_{12} \right\}\ , \label{eq:B25}
\end{equation}
in complete agreement with the equation (\ref{eq:3.72}) derived in the
main text.

\section{A compendium of formulas for moments}
\label{sec:apc}

We list below the expressions of the time-derivatives of moments which
are used in the computation of the waveform and energy loss. For the
waveform:
\begin{mathletters}
\label{eq:C1}
\begin{eqnarray}
{I}^{(2)}_{ij} &=&2\nu m\,{\rm STF}_{ij}
\left\{ v^{ij} - \frac{Gm}{r^3} x^{ij}
+ \gamma\; \frac{Gm}{r^3}\;\;\frac{x^{ij}}{42}\;\;( 149 - 69\nu ) -
\frac{\gamma}{42}v^{ij}\;\;( 23 - 27\nu ) \right.\nonumber \\
&&+ \frac{\gamma^2}{1512}\;\;x^{ij}\;\;\frac{Gm}{r^3} ( -7043 + 7837\nu -
3703\nu^2 )\nonumber \\
&&\left.+\frac{\gamma^2}{1512}\;\;v^{ij}\;\;(-4513-19591\nu+1219\nu^2)\right\}
   \label{eq:C1a}\\
{I}^{(3)}_{ijk} &=& \nu m ( X_2 - X_1 )\,
 {\rm STF}_{ijk} \left[
6v^{ijk} - 21\;\;\frac{Gm}{r^3}\;\;x^{ij} v^k  \right. \nonumber\\
&&\left. - \gamma ( 7 - 8\nu ) v^{ijk} +(83-40\nu)\gamma\;\;\frac{Gm}{r^3}
\;\;x^{ij}v^k \right]\ , \label{eq:C1b}\\
{I}^{(4)}_{ijkl} &=& \nu m\, {\rm STF}_{ijkl}
  \left\{ 24(1-3\nu) v^{ijkl}
  - 192(1-3\nu) \frac{Gm}{r^3}\;\;x^{ij} v^{kl}+
 40(1-3\nu)\;\;\frac{G^2m^2}{r^6}\;\;x^{ijkl}\;\;\right.\nonumber\\
&&+ {288\over 55}\,\gamma
\frac{Gm}{r^3}\;\; x^{ij} v^{kl}\;(161 - 585\nu + 255\nu^2)\nonumber \\
&&+ {4\over 55}\,
\gamma \frac{G^2m^2}{r^6}\; x^{ijkl}\;(-3909 + 13495\nu
- 4695\nu^2) \nonumber \\
&&+\left. {12\over 11} \gamma\;v^{ijkl}\;(-41+183\nu-139\nu^2)\right\}
   \ ,  \label{eq:C1c}\\
{I}^{(5)}_{ijklm} &=& 5\nu m (1 - 2\nu) (X_2 - X_1) \,
 {\rm STF}_{ijklm} \biggl[ 24 v^{ijklm} \nonumber \\
&&\left.- 360\;\;\frac{Gm}{r^3}\;\;x^{ij} v^{klm} + 241
\frac{G^2m^2}{r^6}\;\;x^{ijkl} v^m \right]\ , \label{eq:C1d} \\
{I}^{(6)}_{ijklmn} &=& 24\nu m (1 - 5\nu + 5\nu^2)\,
{\rm STF}_{ijklmn} \biggl[ 30 v^{ijklmn} \nonumber \\
&& \left.- 750 \frac{Gm}{r^3}\;\;x^{ij} v^{klmn} + 1070
\frac{G^2m^2}{r^6}\;\;x^{ijkl} v^{mn}
- 94 \frac{G^3m^3}{r^9}\;\;x^{ijklmn} \right]\ , \label{eq:C1e}\\
{J}^{(2)}_{ij} &=&- \nu m (X_2 - X_1)\,\frac{Gm}{r^3}\,
{\rm STF}_{ij}\,\varepsilon_{jab} x^{ai} v^b \left\{ 1-\;\;
\frac{\gamma}{28}\;(17-20\nu)\right\}\ , \label{eq:C1f} \\
{J}^{(3)}_{ijk} &=&- 8\nu m \frac{Gm}{r^3}\,
{\rm STF}_{ijk}\,
\varepsilon_{kab} x^{ai} v^{bj} \left[ 1-3\nu+
\frac{\gamma}{90}\;\;(-103 + 425\nu - 275\nu^2)\right]\ ,\label{eq:C1g} \\
{J}^{(4)}_{ijkl} &=& 3\nu m (1 - 2\nu) (X_2 - X_1)\;\;
\frac{Gm}{r^3}\;\;{\rm STF}_{ijkl} \left\{
 \varepsilon_{lab} x^av^b (-20 x^iv^{jk}+7\;\;\frac{Gm}{r^3}\;x^{ijk})
\right\} \ , \label{eq:C1h}\\
{J}^{(5)}_{ijklm}&=&32\nu m (1-5\nu+5\nu^2)\;\;\frac{Gm}{r^3}\;\;
{\rm STF}_{ijklm} \left\{ 
\varepsilon_{mab} x^av^b (-15 x^iv^{jkl}+17\;\frac{Gm}{r^3}\;\;x^{ijk}
v^l) \right\}\ . \label{eq:C1i}
\end{eqnarray}
\end{mathletters}
For the energy loss:
\begin{mathletters}
\label{eq:C2}
\begin{eqnarray}
{I}^{(3)}_{ij}&=& -8\nu m\, {\rm STF}_{ij}{Gm
\over r^3}v^ix^j\left\{1\;-\;\frac{\gamma}{42}(149-69\nu)\right.\nonumber\\
&&\left.\;+\;\frac{\gamma^2}{1512}(7043-7837\nu+3703\nu^2)\right\} \ ,
\label{eq:C2a}\\
{I}^{(4)}_{ijk}&=&\nu m(X_2-X_1)\, {\rm STF}_{ijk}\frac{Gm}{r^3}
\left\{\left[ 21-\gamma(146-61\nu)\right]x^{ijk}
\frac{Gm}{r^3}\right.\nonumber\\
&&\;-\;[60-\gamma(241-122\nu)]v^{ij}x^k\biggr\}\ ,\label{eq:C2b}\\
{I}^{(5)}_{ijkl}&=&-8\nu m(1-3\nu)\, {\rm STF}_{ijkl}
\frac{Gm}{r^3}\left\{60v^{ijk}x^l
-68\frac{Gm}{r^3}x^{ijk}{v^l}\right\} \ ,\label{eq:C2c}\\
{J}^{(3)}_{ij}&=& \nu m(X_2-X_1)\, {\rm STF}_{ij}
\frac{Gm}{r^3}\left\{ -1+\frac{\gamma}{28}(17-20\nu)\right\}
\varepsilon^{jab}x^av^{bi}\ ,\label{eq:C2d}\\
{J}^{(4)}_{ijk}&=& -8\nu m (1-3 \nu)\,{\rm STF}_{ijk}
  \frac{Gm}{r^3}\varepsilon^{iab}
\left\{x ^av^{bjk}-\frac{Gm}{r^3}x^{ajk}v^b\right\}\ . \label{eq:C2e}
\end{eqnarray}
\end{mathletters}
The TT projections of relevant contractions of time-derivatives of moments
with {\bf N} as used in the waveform are:
\begin{mathletters}
   \label{eq:C3}
\begin{eqnarray}
{\cal P}_{ijkm}{I}^{(2)}_{ij} &=& 2\nu m {\cal P}_{ijkm}
\left\{ v^{ij} -\; \frac{Gm}{r}\;\;n^{ij}\right.\nonumber \\
&&+ \frac{\gamma}{42}\;\;\frac{Gm}{r}\;\;n^{ij} (149 - 69\nu)\;\;
-\;\;\frac{\gamma}{42}\;\; (23 - 27\nu) v^{ij} \nonumber \\
&&+ \frac{\gamma^2}{1512}\;\;\frac{Gm}{r}\;\;n^{ij} (-7043 + 7837\nu -
3703\nu^2) \nonumber \\
&&+ \left.\frac{\gamma^2}{1512} v^{ij}(-4513 - 19591\nu + 1219\nu^2)\right\}
\ , \label{eq:C3a}\\
{\cal P}_{ijkm}{I}^{(3)}_{ija}N_a &=& \nu m (X_2 - X_1)
 {\cal P}_{ijkm}\left\{ 6(vN)
v^{ij} - 7(vN) \frac{Gm}{r} n^{ij}\right. \nonumber \\
&&- 14(nN) \frac{Gm}{r} n^{i} v^{j} - \gamma (vN) (7-8\nu) v^{ij}\nonumber \\
&&+\left. \frac{1}{3}(83-40\nu)\gamma (vN) \frac{Gm}{r} n^{ij} +
\frac{2}{3}(83-40\nu)\gamma (nN) \frac{Gm}{r} n^{i}
v^{j}\right\} \ , \label{eq:C3b}\\
{\cal P}_{ijkm}{I}^{(4)}_{ijab}\,\,N_{ab} &=& \nu m
{\cal P}_{ijkm} \left\{ (1-3\nu)\left[24(vN)^2 v^{ij} +
\frac{8}{7}\,\,\frac{Gm}{r}\,\,v^{ij}\right.\right.\nonumber \\
&&- 32\,\frac{Gm}{r}\,\,(nN)^2 v^{ij} -
32(vN)^2\,\,\frac{Gm}{r}\,\,n^{ij}\,\,-\,\frac{8}{7}\,\,
\frac{G^2m^2}{r^2}\,\,n^{ij}\nonumber \\
&&\left.+ 40(nN)^2\,\,\frac{G^2m^2}{r^2}\,\,n^{ij} - 128(nN)(vN)\,\,
\frac{Gm}{r}\,\,n^{i}v^{j} \right] \nonumber \\
&&+ \gamma \left[\frac{12}{385}\,\,(-109 + 325\nu + 5\nu^2)\,\,
\frac{Gm}{r}\,\,v^{ij}\right. \nonumber \\
&&+ \frac{12}{11}\,\,(-41 + 183\nu - 139\nu^2) (vN)^2 v^{ij} \nonumber \\
&&+ \frac{48}{55}\,\,(161 - 585\nu + 255\nu^2)
(nN)^2\,\,\frac{Gm}{r}\,\,v^{ij} \nonumber \\
&&+ \frac{4}{385}\,\,(657 - 2075\nu +
315\nu^2)\,\,\frac{G^2m^2}{r^2}\,\,n^{ij} \nonumber \\
&&+ \frac{48}{55}\,\,(161 - 585\nu + 255\nu^2)\,\,\frac{Gm}{r}
\,(vN)^2 n^{ij} \nonumber \\
&&+ \frac{4}{55}\,\,(-3909 + 13495\nu - 4695\nu^2)
(nN)^2\,\,\frac{G^2m^2}{r^2}\,\, n^{ij}\nonumber \\
&&+\left.\left. \frac{192}{55}\,\,(161 - 585\nu + 255\nu^2) (nN)
(vN)\,\,\frac{Gm}{r}\,\, n^{i}v^{j} \right] \right\}\ , \label{eq:C3c}\\
{\cal P}_{ijkm}{I}^{(5)}_{ijabc}\,\,N_{abc} &=& 5\nu m (1-2\nu)
(X_2-X_1){\cal P}_{ijkm}\left[24\left\{(Nv)^3 v^{ij}-\frac{v^2}{3}(Nv) v^{ij}
  \right\}\right. \nonumber \\
&&- 36\,\frac{Gm}{r} \biggl\{ n^{ij} (Nv)^3 + 6n^{i} v^{j} (Nn) (Nv)^2 +
  3(Nn)^2(Nv) v^{ij} \nonumber \\
&&-\left.\frac{1}{3}\,(Nv) v^{ij}-{1\over 3} v^2 (Nv) n^{ij}
   - {2\over 3}(Nn) v^2 n^{i} v^{j} \right\}\nonumber \\
&&+ \frac{241}{5}\,\,\frac{G^2m^2}{r^2}\,\,\biggl\{ 3n^{ij} (Nn)^2 (Nv) +
2n^{i} v^{j} (Nn)^3\nonumber \\
&&-\left.\left. \frac{1}{3}\,\,n^{ij} (Nv) -{2\over 3} n^{i} v^{j} (Nn)
         \right\}\right]\ , \label{eq:C3d} \\
{\cal P}_{ijkm}\,{I}^{(6)}_{ijabcd}\,\,N_{abcd}&=&24\nu m(1-5\nu +
    5\nu^2){\cal P}_{ijkm}\biggl[
    30 \left\{v^{ij}(Nv)^4 - \frac{6}{11}\,v^2v^{ij}(Nv)^2 +
    \frac{1}{33}\,v^4v^{ij}\right\} \nonumber \\
&&-\,50\,\frac{Gm}{r}  \biggl\{ n^{ij}(Nv)^4 +
    8n^{i}v^{j}(Nn)(Nv)^3 + 6v^{ij}(Nn)^2(Nv)^2\nonumber \\
&&-\frac{6}{11}\,[v^2 (n^{ij}(Nv)^2 + 4n^{i}v^{j}(Nn)(Nv) +
    v^{ij}(Nn)^2) + v^{ij}(Nv)^2]\nonumber \\
&&+\left.\frac{1}{33}\,(2v^2v^{ij} + v^4n^{ij})\right\}\nonumber \\
&&+\,\frac{214}{3}\,\frac{G^2m^2}{r^2} \biggl\{  6n^{ij}(Nn)^2(Nv)^2 +
    (Nn)^4 v^{ij} + 8n^{i}v^{j}(Nn)^3(Nv)\nonumber \\
&&-\,\frac{6}{11}\,[n^{ij} (Nv)^2 + 4n^{i}v^{j}(Nn)(Nv) + (Nn)^2 v^{ij}
    + v^2n^{ij}(Nn)^2]\nonumber \\
&&+\left.\,\frac{1}{33}\,(v^{ij} + 2v^2 n^{ij})\right\}\nonumber \\
&&-\left.\,94\,\frac{G^3m^3}{r^3}\left[(nN)^4n^{ij} - \frac{6}{11}(Nn)^2n^{ij}
+
\frac{1}{33}\,n^{ij}\right]\right]\ ,  \label{eq:C3e}  \\
  {\cal P}_{ijkm}\varepsilon_{abi}{J}^{(2)}_{ja}\,\,N_b &=&
  -\,\nu m\,(X_2 - X_1)\,\frac{Gm}{r}\,{\cal P}_{ijkm}
  (1-\,\frac{\gamma}{28}\ (17 - 20\nu))\nonumber \\
&&\times [(nN) v^{i}n^{j} - (vN) n^{ij}]\ , \label{eq:C3f}\\
{\cal P}_{ijkm}\varepsilon_{abi}\,{J}^{(3)}_{jac}\,\,N_{bc} &=&
  -\,\frac{4}{3}\,\nu m\,\frac{Gm}{r}\,{\cal P}_{ijkm}\left\{ (1-3\nu)
  \left[\,\frac{Gm}{r}\,n^{ij} - v^{ij} + 3 (nN)^2 v^{ij}
  - 3(vN)^2 n^{ij}\right]\right. \nonumber \\
&&+ \frac{\gamma}{90}\,\left[\frac{Gm}{r}\, n^{ij} (-373 + 1325\nu -
  545\nu^2 )\right. \nonumber \\
&&+ (3(nN)^2 v^{ij} -3(vN)^2 n^{ij} - v^{ij})
  (-103 + 425\nu - 275\nu^2)\biggr] \biggr\}\ , \label{eq:C3g}  \\
{\cal P}_{ijkm}\varepsilon_{abi}\,{J}^{(4)}_{jacd}N_{bcd}
&=& 3\nu m (1-2\nu) (X_2 - X_1) \frac{Gm}{r}\,{\cal P}_{ijkm}\nonumber \\
&&\times\left[-\frac{5}{3} \biggl\{-4v^{i}n^{j} (Nv)^2 (Nn)
  + 8v^{ij} (Nn)^2 (Nv)\right.\nonumber \\
&&+\,2(Nv)v^2 n^{ij} + 2(Nn)v^2 n^{i}v^{j} - 4n^{ij}(Nv)^3 -
   2v^{ij}(Nv)\nonumber \\
&&-\left.\,\frac{v^2}{7} (2v^{i}n^{j} (Nn) - 2n^{ij} (Nv))\right\}\nonumber \\
&&+\,\frac{7}{4}\,\frac{Gm}{r}\,\biggl\{4(Nn)^3 v^{i}n^{j} - 4(Nv) (Nn)^2
  n^{ij}\nonumber \\
&&-\left.\left.\,2(Nn) v^{i}n^{j} -\,\frac{2}{7} ((Nn)v^{i}n^{j} -
     (Nv)n^{ij})\right\}\right] \ , \label{eq:C3h}\\
{\cal P}_{ijkm}\varepsilon_{abi}\,{J}^{(5)}_{jacde}N_{bcde} &=&
    32\nu m (1
    - 5\nu + 5\nu^2)\,\frac{Gm}{r}{\cal P}_{ijkm}\nonumber \\
&&\times \left[-\,\frac{15}{20}\,\biggl\{-10(Nv)^3(Nn)v^{i}n^{j} +
   6v^2(Nn)(Nv)n^{i}v^{j}\right.\nonumber \\
&&+\,4v^2(Nv)^2n^{ij} - 5(Nv)^4n^{ij} + 15(Nn)^2(Nv)^2v^{ij}\nonumber \\
&&-\left.\,3(Nv)^2v^{ij} - v^2(Nn)^2 v^{ij} -{1\over 3}v^4n^{ij}
  + {1\over 3} v^2 v^{ij}\right\}\nonumber \\
&&+\,\frac{17}{20}\,\frac{Gm}{r}\,\biggl\{10(Nn)^3(Nv)v^{i}n^{j} -
  6(Nn)(Nv)v^{i}n^{j}\nonumber \\
&&-\,15(Nn)^2(Nv)^2n^{ij} + 5(Nn)^4v^{ij}
  -\,4(Nn)^2v^{ij} + 3v^2(Nn)^2n^{ij}\nonumber \\
&&\left.\left.\, +(Nv)^2n^{ij}
  -{1\over 3} v^2n^{ij} +{1\over 3}
  v^{ij}\right\}\right]\ .  \label{eq:C3i}
\end{eqnarray}
\end{mathletters}

Finally the ``squares'' of time-derivatives of moments as used in the
energy loss are:
\begin{mathletters}
\label{eq:C4}
\begin{eqnarray}
({I}^{(3)}_{ij})^2&=&32(\nu m)^2\frac{G^3m^3}{r^5}\left\{1\;-\;
\frac{\gamma}{21}(212-90\nu)\right.\nonumber\\
&&\;\left.+\;\frac{\gamma^2}{2646}(130150-76007\nu+31442\nu^2)\right\}
\ ,\label{eq:C4a}\\
({I}^{(4)}_{ijk})^2&=&(\nu m)^2\frac{G^2m^2}{r^4}
\frac{(1-4\nu)}{15}\left\{
126\frac{G^2m^2}{r^2}[21-2\gamma(146-61\nu)] \right.\;\nonumber\\
&&\;+\;480v^4[30-\gamma(241-122\nu)] \nonumber\\
&&\left.\;+\;18\frac{Gm}{r}v^2[420-\gamma(4607-2074\nu)]\right\}\ ,
\label{eq:C4b}\\
({I}^{(5)}_{ijkl})^2&=&\frac{512}{7}(\nu m)^2(1-3\nu)^2
\frac{G^2m^2}{r^4}
\left\{450 v^6+578 \frac{G^2m^2}{r^2}v^2
+765\frac{Gm}{r}v^4\right\}\ , \label{eq:C4c}\\
({J}^{(3)}_{ij})^2&=&(\nu m)^2\frac{G^2m^2}{r^4}v^4
\frac{(1-4\nu)}{2}
\left\{1-\frac{\gamma}{14}(17-20\nu)\right\}\ ,\label{eq:C4d}\\
({J}^{(4)}_{ijk})^2&=&\frac{128}{15}(\nu m)^2(1-3\nu)^2
\frac{G^2m^2}{r^4}\left\{
2v^6+2\frac{G^2m^2}{r^2}v^2+\frac{Gm}{r}v^4\right\}\ . \label{eq:C4e}
\end{eqnarray}
\end{mathletters}


\begin{references}
\bibitem{N1}In the case,
considered here, of quasi-circular orbits it is sufficient to consider
the energy loss. This case brings also the simplification of anulling
any possible periodic difference, in the form of a total time
derivative, between the local 2PN mechanical energy loss and the
radiated energy flux at infinity.
\bibitem{N2}For simplicity, we are discussing here the case of extended
bodies where $U({\bf x})$ and $W_{ij}({\bf x})$ are everywhere regular.
Even in the point-mass limit it is enough that $W({\bf x})$ be regular
in the generic case where the origin ${\bf x}=0$ differs from the
positions of the point masses ${\bf y}_A$.
\bibitem{N3}Note that, in the limit $\nu =0$, the parameter $x$ reduces
to the Schwarzschild-coordinate version of the parameter $\gamma$:
$\gamma_{\rm Sch.}=Gm/(r_{\rm Sch.}c^2)$ where $r_{\rm Sch.}=r+Gm/c^2$.
Ref.~\cite{TSasa94} denotes the parameter $x=(Gm\omega /c^3)^{2/3}$ by
$v^2$.
\bibitem{N4}The result (\ref{eq:4.24}) agrees with Eq.~(44) in
Ref.~\cite{SW93}.  Note that the latter reference uses ADM coordinates
instead of harmonic coordinates.  In particular the ADM-coordinate
radius of the circular orbit is related to the harmonic-coordinate one
by $r_{\rm ADM} = r [1-{1\over 8} (2+29\nu) \gamma^2]$.
\end{references}
\end{document}